\documentclass[11pt, a4paper, reqno]{amsart}
\usepackage{color}
\usepackage{amsmath}
\usepackage{amsthm}
\usepackage{amssymb}
\usepackage{amsmath}
\usepackage{mathcomp}
\usepackage{dsfont}
\usepackage[toc,page]{appendix}
\usepackage{graphicx}
\usepackage{subfigure}
\usepackage{enumerate}
\usepackage{amsfonts}
\usepackage{amsthm}
\usepackage{amsmath}
\usepackage{times}
\usepackage{amscd}
\usepackage{t1enc}
\usepackage{enumitem}
\usepackage[mathscr]{eucal}
\usepackage{indentfirst}
\usepackage{graphicx}
\usepackage{graphics}
\usepackage{pict2e}
\usepackage{epic}
\usepackage{esint}
\usepackage[margin=2.9cm]{geometry}
\usepackage{epstopdf}
\usepackage{verbatim}
\usepackage{cancel}
\usepackage{amssymb}
\usepackage{xcolor}
\usepackage{dsfont}
\usepackage{hyperref}
\usepackage{physics}
\usepackage{amsmath}

\usepackage{mathtools}
\mathtoolsset{showonlyrefs} %to hide unquoted eqs

\usepackage[margin=2.9cm]{geometry}

\newcommand{\bE}{\mathbb E}
\newcommand{\R}{\mathbb{R}}
\newcommand{\N}{\mathbb{N}}
\newcommand{\cP}{\mathcal{P}}
\newcommand{\suchthat}{\ensuremath{\ : \ }} % such that inside, for example, the sets definition
\newcommand{\ler}[1]{\left( #1 \right)}
\newcommand{\lers}[1]{\left\{ #1 \right \}}
\newcommand{\cW}{\mathcal{W}}
\newcommand{\cC}{\mathcal{C}}

\usepackage{hyperref}

\setitemize{itemsep=5pt, topsep=7pt}
\numberwithin{equation}{section}

\newtheorem{definition}{Definition}[section]
\newtheorem{lemma}[definition]{Lemma}

\newtheorem{theorem}[definition]{Theorem}

\numberwithin{equation}{section}

\def\cN{\mathcal{N}}
\def\a0{{\rm a}_0}

\def\tr{\mathrm{tr}}

\def\law{\mathrm{law}}

\newcommand{\vertiii}[1]{{\left\vert\kern-0.25ex\left\vert\kern-0.25ex\left\vert #1
    \right\vert\kern-0.25ex\right\vert\kern-0.25ex\right\vert}}

\title{Limit theorems for empirical measures of interacting quantum systems in Wasserstein space}
\begin{document}

\author{
Lorenzo Portinale ,
Simone Rademacher, 
and 
D\'aniel Virosztek }

\address{Lorenzo Portinale, Hausdorff Center for Mathematics, Institut f\"ur Angewandte Mathematik, Endenicher Allee 60, 53115 Bonn, Germany \texttt{portinale@iam.uni-bonn.de}}
\address{Simone Rademacher, Department of Mathematics, Ludwig-Maximilians-Universit\"at M\"unchen, Theresienstr. 39, 80333 Munich, Germany. \texttt{simone.rademacher@math.lmu.de}}
\address{D\'aniel Virosztek, Alfr\'ed R\'enyi Institute of Mathematics, Re\'altanoda street 13-15, H-1053, Budapest, Hungary. \texttt{virosztek.daniel@renyi.hu}}

\date{\today}

\begin{abstract}
We prove fundamental properties of empirical measures induced by measurements performed on quantum $N$-body systems. More precisely, we consider measurements performed on the ground state of an interacting Bose gas on the torus in the Gross--Pitaevskii regime, known to exhibit Bose--Einstein condensation. For the corresponding empirical measure, we prove a weak law of large numbers with limit induced by the condensate wave function and characterize the fluctuations around through an appropriate central limit theorem. 
\end{abstract}

\maketitle

\section{Introduction}

Recent decades have seen a great deal of progress in the study of random empirical measures induced by i.i.d. random variables \cite{AKT, simple-AKT, Bobkov-Ledoux, founier-guillin}. If $\mu_N$ is the empirical measure corresponding to an $N$-sample of the law $\mu \in \mathrm{Prob} (\R^n),$ that is, $\mu=\frac{1}{N}\sum_{j=1}^N \delta_{X_j}$ where $X_1, \dots, X_N$ are i.i.d. and $\mathrm{law}(X_1)=\mu,$ then the well-known Glivenko--Cantelli theorem tells us that the weak limit of $\mu_N$ is $\mu$ almost surely. So it is not the identification of the limit but the study of the rate of convergence which gained a lot of attention in the past years. A quantitative study of the rate of convergence requires a satisfactory notion of distance on probability measures. The $p$-Wasserstein distances, which are defined by optimal transportation (see Subsection \ref{susec:OT-intro} for precise definitions), have proved to be particularly useful in quantifying the dissimilarity of an empirical measure and its background distribution \cite{Bobkov-Ledoux, founier-guillin}. It should be noted that this problem is highly non-trivial even in one dimension --- as it can be seen from the work of Bobkov and Ledoux \cite{Bobkov-Ledoux}. Another important note is that the problem of \emph{optimal matchings} pioneered by Ajtai--Koml\'os--Tusn\'ady \cite{AKT,simple-AKT} is intimately related to this problem: the optimal matching problem concerns the $p$-Wasserstein distance of empirical measures corresponding to independent $N$-samples of the same background law.  
\par
On the other hand, there have been a great interest recently in studying random variables arising in the context of quantum many-body systems. These works focused on the \emph{average} of the outcomes of measurements performed on every single particle. It is a feature of the actual physical setting that the measurements of a one-particle observable on different particles give rise to \emph{mildly dependent} random variables. Despite this lack of independence, the behaviour of this average has been successfully studied in several interesting physical situations. The fluctuation of this average around its expectation has been characterized by central limit theorems \cite{RS,R-2020}, and these results have been complemented by large deviation estimates providing effective bounds on probabilities of outliers \cite{R, R-LD2023, RS-JSP-22}.
\par
We aim to merge the above two research directions in the following sense: we are studying the fundamental properties of \emph{empirical measures induced by measurements performed on quantum $N$-body} systems. As the measurement of a fixed one-particle observable on each particle simultaneously gives rise to a \emph{dependent} ensemble of random variables, our first task is to determine the limit of the empirical measures. Our corresponding result is a law of large numbers (Theorem \ref{thm:lln}) saying that the limit is the law induced by the condensate wave function. Our second main result concerns the fluctuations of the empirical measures around their limit: we were able to prove an appropriate central limit theorem in the infinite-dimensional space of signed Borel measures on the real line --- see Theorem \ref{thm:clt}. It is the main goal of a future work of ours to complement these results with appropriate large deviation estimates.

\subsection*{Acknowledgments}
L.P. gratefully acknowledges fundings from the Deutsche Forschungsgemeinschaft (DFG, German Research Foundation) under Germany's Excellence Strategy - GZ 2047/1, Projekt-ID 390685813.  Financial support by the Deutsche Forschungsgemeinschaft (DFG) within the CRC 1060, at University of Bonn project number 211504053, is also gratefully acknowledged.
%
% S.R.
%
D.V. was supported by the Momentum Program of the Hungarian Academy of Sciences (grant no. LP2021-15/2021) and partially supported by the ERC Consolidator Grant no. 772466. This project was initiated during the thematic semester "Optimal Transport on Quantum Structures" held at the Erd\H{o}s Center, Budapest in the fall of 2022.
\section{Results}

Before stating our main results in Theorem \ref{thm:lln} and Theorem \ref{thm:clt} we will first introduce the interacting Bose gas, the $N$-body quantum system this paper is dedicated to, and then define $p$- Wasserstein metrics, the mathematical object we use to analyse the Bose gas.

\subsection{An introduction to the interacting Bose gas.} Physically, trapped Bose gases of a large number $N$ of particles are of peculiar interest as they undergo a phase transition at extremely low temperatures: the majority, i.e. $O(N)$,  of the bosons condenses into the same quantum state, which is called Bose-Einstein condensate. This phenomenon was first predicted theoretically by Bose \cite{Bose} and Einstein \cite{Einstein} in 1924 and later confirmed experimentally by working groups of Wiemann \cite{Wiemann} and Ketterle \cite{Ketterle}, for which the nobel prize was awarded in 2001.  
Since then, the mathematical description and prediction of Bose-Einstein condensation, that is based on the basic principles of quantum mechanics, has been one of the major problems of mathematical physics. 

Mathematically, a Bose gas of $N$ interacting particles is described by the basic principles of quantum mechanics, namely many-body theory, through a normalized wave function in the Hilbert space $L^2_s ( \Lambda^N)$, that is the symmetric subspace of $L^2( \Lambda^N)$. In the following we assume the $N$ bosons being trapped on the unit torus in three dimensions $\Lambda =[0;1]^3$ and described through the Hamiltonian 
\begin{align}
\label{eq:ham}
H_N = \sum_{i=1}^N (-\Delta_j) + \frac{1}{N} \sum_{i,j =1}^N V_N (x-y ) 
    \, , 
\end{align} 
with scaled two-body interaction potential $V_N = N^3 V ( N \cdot )$. The interaction potential's scaling with the number of particles $N$ is called Gross--Pitaevskii scaling and motivated by an approximate delta interaction in the limit of a large number of interacting particles $N \rightarrow \infty$. In the following we consider the system at zero temperature at which it relaxes to the ground state $\psi_N$ of $H_N$.

In order to give a mathematical formulation of Bose--Einstein condensation, we consider a one-particle observable that is a one-particle operator $O$ $L^2(\Lambda) \rightarrow L^2( \Lambda)$. Measuring the one-particle observable $O$ on each of the $N$ bosons simultaneously gives rise to an $\R^N$-valued random variable $(Y_i^N)_{i=1}^N$ having the distribution
\begin{align} 
\label{eq:law-of-y}
\law (Y_1^N, \dots, Y_N^N) ( A_1 \times \dots \times A_N ) =& \mathbb{P} \left( Y_1^N \in A_1, \dots, Y_N^N \in A_N \right) \notag \\
=& \langle \psi_N, \, E(A_1) \otimes E(A_2) \otimes \dots \otimes E(A_N) \psi_N \rangle 
    \, , 
\end{align}
where $\psi_N$ denotes the ground state of $H_N$ given by \eqref{eq:ham}, whereas  $E$ denotes the spectral measure of the observable $O$, that is the projection valued measure $E: \mathcal{B}(\R) \rightarrow \mathcal{P}(L^2(\Lambda))$ characterized by $O=\int_{\R} \lambda \dd E(\lambda)$. Here $\mathcal{B}(\R)$ stands for the Borel $\sigma$-algebra on $\R,$ and $\mathcal{P}(L^2(\Lambda))$ denotes the projection lattice on $L^2(\Lambda)$.
Now let the $N$-particle operators $O^{(i)}$ with $i \in \{ 1, \dots, N \}$ be defined by
\begin{align} \label{eq:o-i-def}
O^{(i)} = \mathds{1} \otimes \cdots \otimes \mathds{1} \otimes O \otimes \mathds{1} \otimes \cdots \otimes \mathds{1} \, ,
\end{align}
that is, $O^{(i)}$ is the operator acting as $O$ on the $i$-th particle and as identity on the remaining $N-1$ particles.
By the above definition \eqref{eq:o-i-def}, $\mathds{1}_{A_i}(O^{(i)})$ commutes with $\mathds{1}_{A_j}(O^{(j)})$ for any $1 \leq i,j \leq N,$ and 
\begin{equation} \label{eq:1-E-corresp}
\mathds{1}_{A_1}(O^{(1)}) \dots \mathds{1}_{A_N} (O^{(N)})=E(A_1) \otimes E(A_2) \otimes \dots \otimes E(A_N).
\end{equation}
Consequently,
$$
\mathbb{P} \left( Y_1^N \in A_1, \dots, Y_N^N \in A_N \right)
=\langle \psi_N, \, \mathds{1}_{A_1}(O^{(1)}) \dots \mathds{1}_{A_N} (O^{(N)}) \psi_N \rangle.
$$
Note that with this definition we have
\begin{align}
\bE \left[ f(Y_i^N) \right] = \int_{\mathbb{R}} f(y) \, d \law (Y_i^N) (y)  = \langle \psi_N, f(O^{(i)}) \psi_N \rangle = \tr [\gamma_{N}^{(1)} f(O)] \, , 
\label{eq:expec}
\end{align}
for any Borel measurable real function $f,$ and for the one particle reduced density $\gamma_N^{(1)}$ defined by 
\begin{align}
\label{def:gamma}
\gamma_N^{(1)} = \tr_{2, \dots, N} \vert \psi_N \rangle \langle \psi_N \vert \, . 
\end{align}
In fact the last equality in \eqref{eq:expec} follows by the symmetry of the ground state $\psi_N$ and shows, in particular, that the random variables $(Y_i^N)_{i =1}^N$ are identically distributed. Moreover $(Y_i^N)_{i =1}^N$ are not independent since 
\begin{align}
\mathbb{P} \left( Y_i^N \in A_i, Y_j^N \in A_j \right)
= \langle \psi_N, \, \mathds{1}_{A_i}(O^{(i)}) \mathds{1}_{A_j} (O^{(j)}) \psi_N \rangle = \tr \gamma_N^{(2)} \left( \mathds{1}_{A_i}(O) \otimes \mathds{1}_{A_j} (O)  \right)
\end{align}
and $\psi_N \not= \varphi^{\otimes N}$, i.e. $\gamma_N^{(2)} \not= \vert \varphi \rangle \langle \varphi \vert^{\otimes 2}$, due to correlations of the interacting particles (see \cite{LS_cond} for a first resp. \cite{BBCS_cond,BBCS_optimal,BBCS,NR} for a more refined analysis of the correlation structure of the particles).

%As a consequence of Bose-Einstein condensation, the random variables $(Y_i^N)_{i=1}^N$, first introduced in \cite{BKS},  

The goal of this paper is to compare the empirical measure $\nu_N : ( \Omega, \mathbb{P}) \rightarrow \mathcal{P} ( \mathbb{R})$ given by
 \begin{align}
 \label{def:nu_N}
  \nu_N := \frac{1}{N} \sum_{i=1}^N \delta_{Y_i^N} \, , 
 \end{align}
where the random variables $(Y_i^N)_{i=1}^N$ are defined by \eqref{eq:law-of-y}, with the probability measure $\nu_{\varphi}$ that corresponds to the condensate wave function $\varphi= 1_{\Lambda}$ and is given by 
\begin{align}
\label{def:nu-phi}
\nu_{\varphi}(A)=\langle \varphi, \mathds{1}_A(O) \varphi \rangle \, ,
\end{align}
for any Borel set $A \subset \R$.

\subsection{An introduction to optimal transport and Wasserstein distances} \label{susec:OT-intro}

As our main findings (Theorem \ref{thm:lln} and Theorem \ref{thm:clt}) are law of large numbers and central limit theorem type results for empirical measures, we need a good notion of distance on probability measures to quantify the dissimilarity of an actual (random) empirical measure from its reference measure. A highly reasonable choice of such a distance is the \emph{Wasserstein distance} defined by optimal transportation.  
\par
Given a complete and separable metric space $(X, \rho)$ and a parameter $p \geq 1,$ the \emph{$p$-Wasserstein space over $X$} is denoted by $\mathcal{W}_p(X)$ and is defined to be the collection of all Borel probability measures on $(X,\rho)$ with finite moment of order $p.$ That is,
\begin{equation} \label{eq:wass-space-def}
    \mathcal{W}_p(X)=\lers{\mu \in \mathrm{Prob}(X) \, \middle| \, \int_{X} \rho(x,x_0)^p \dd \mu(x) < \infty \text{ for some } x_0 \in X }.
\end{equation}
A \emph{transport plan} between probability measures $\mu, \nu \in \mathcal{W}_p(X)$ is a probability measure $\pi$ on the product space $X \times X$ such that the marginals of $\pi$ are $\mu$ and $\nu,$ that is,
\begin{equation} \label{eq:coupling-def}
\pi(A \times X)=\mu(A) 
    \quad    \text{and}     \quad  
\pi(X \times B)=\nu(B), 
    \quad \text{for all } A,B \subset X \text{ Borel sets}.    
\end{equation}
We denote by $\mathcal{C}(\mu,\nu)$ the set of all $\pi'$s that satisfy \eqref{eq:coupling-def}. If the cost of moving one unit of mass from $x$ to $y$ is described by the cost function $c: X \times X \rightarrow \R; \, (x,y) \mapsto c(x,y),$ then the cost of a transport plan $\pi \in \mathcal{C}(\mu,\nu)$ is given by $\iint_{X^2} c(x,y) \dd \pi(x,y).$ Now the $p$-Wasserstein distance of $\mu$ and $\nu$ corresponds to the optimal cost of transforming $\mu$ into $\nu$ when the cost function is chosen to be the $\rho^p$. That is,
\begin{equation} \label{eq:wass-dist-def}
d_{\cW_{p}}(\mu,\nu)=\ler{\inf\lers{\iint_{X^2} \rho(x,y)^p \dd \pi(x,y) \, \middle| \, \pi \in \cC(\mu,\nu)}}^\frac{1}{p}.
\end{equation}
An important feature of the $p$-Wasserstein distance is that is metrizes the weak topology if $X$ is a bounded metric space, and it defines a topology slightly stronger than the weak for general (unbounded) $X$ --- see, e.g., \cite[Theorem 7.12]{V}. For more details and references on optimal transport theory and Wasserstein spaces we refer the reader to any of the following comprehensive textbooks \cite{AG,Figalli, Santambrogio,Villani,V}. 

\subsection{Main results} Now we are ready to state our main results. The first result is about a law of large numbers:  
 
\begin{theorem}
\label{thm:lln}
Let $V \in L^3 ( \R^3)$ be compactly supported, non-negative and spherically symmetric and $\psi_N$ denote the (uniquely) defined ground state of $H_N$ given by \eqref{eq:ham}. For a bounded operator $O$ on $L^2( \Lambda )$, let $\nu_N$ denote the empirical measure given by \eqref{def:nu_N} and $\nu_\varphi$ by \eqref{def:nu-phi}. Let $\delta >0$, then
\begin{align}
\lim_{N \rightarrow \infty}\mathbb{P} \left[ \vert W_1 ( \nu_N, \nu_\varphi ) \vert  > \delta \right]   =0 \, .
\end{align}
\end{theorem}

To our knowledge Theorem \ref{thm:lln} provides a first characterization of the empirical measure $\nu_N$ and thus a novel approach to the description of Bose-Einstein condensation.  \\

\paragraph{\textit{Law of large numbers for $(Y_i^N)_{i=1}^N$}.} We stress that the random variables $(Y_i^N)_{i=1}^N$, first introduced in \cite{BKS}, are well known to satisfy a weak law of large numbers that is an immediate consequence of Bose-Einstein condensation of the ground state $\psi_N$. Bose-Einstein condensation of $\psi_N$ is mathematically defined as trace norm convergence of the one-particle reduced density 
\begin{align}
\label{eq:BEC}
\gamma_N^{(1)} \rightarrow \vert \varphi \rangle \langle \varphi \vert \quad \text{as} \quad  N \rightarrow \infty  \, , 
\end{align}
where $\gamma_N^{(1)}$ is given by \eqref{def:gamma} and $\varphi = 1_\Lambda$ denotes the condensate wave function. Bose-Einstein condensation \eqref{eq:BEC} was first proven in \cite{LS_cond} and later refined by a more detailed analysis of the particles outside the condensate wave function in \cite{BBCS_cond,BBCS_optimal,BBCS,NR}. For the extension to Bose gases trapped in $\mathbb{R}^3$ through an external potential see for example \cite{BSS,LS,NNRT,NRS,NT}. \\

We can further characterize the fluctuations of the empirical measure $\nu_N$ around its limit $\nu_\varphi$ by a central limit theorem. The limiting Gaussian is given in terms of the scattering length $\a0$ of the un-scaled potential $V$ defined through 
\begin{align}
\label{def:a0}
 8 \pi \a0 = \int  \, V(x) f(x) \dd x
    \, , 
\end{align}
with $f$ the solution to the zero-energy scattering equation 
\begin{align}
\left[ - \Delta + \frac{1}{2} V \right] f =0
\end{align}
with boundary conditions $f(x) \rightarrow 1$ as $\vert x \vert \rightarrow \infty$.

In the statement of our main result, we adopt the standard terminology: we say that a sequence of $\R^m$-valued random variable $Z_N$ converge \textit{vaguely in distribution} to a limit random variable $Z$ whenever the laws of $Z_N$ vaguely converge to the law of $Z$, i.e. 
\begin{align}
    \lim_{N \to \infty} 
    \int_{\R^m} G \dd (\law Z_N)
        =
    \int_{\R^m} G \dd (\law Z) 
        \qquad \forall G \in C_c(\R^m)
            \, .
\end{align}
Similarly, we say that $Z_N$ converge \textit{weakly in distribution} to $Z$ whenever the above limit equality holds for every $G \in C_b(\R^m)$.

\begin{theorem}
\label{thm:clt}
Let $V \in L^3 ( \R^3)$ be compactly supported, non-negative and spherically symmetric and $\psi_N$ denote the (uniquely) defined ground state of $H_N$ given by \eqref{eq:ham}. For a bounded operator $O$ on $L^2( \Lambda )$, let $\nu_N$ denote the empirical measure given by \eqref{def:nu_N} and $\nu_\varphi$ by \eqref{def:nu-phi}.
\begin{enumerate}
\item[(i)] There exists $C>0$ such that
\begin{align}
    \sqrt{N} \bE \left[ W_1 ( \nu_N, \nu_\varphi) \right] \leq C \, . \label{eq:E-W1}
\end{align}
\item[(ii)]
% \textcolor{magenta}{This is not yet proven (see section below), we still need to find a way how to apply the results of \cite{RS} (that is formulated w.r.t. to sufficiently nice functions $g$)} 
For every $m \in \mathbb{N}$ and every $f_1, \cdots, f_m \in C_b ( \mathbb{R})$, we have that 
\begin{align}
\label{eq:CLT_main}
    \lbrace \sqrt{N} \left( \langle \nu_N (\omega),  f_j \rangle - \langle \nu_\varphi,  f_j \rangle \right) \rbrace_{j=1}^m \to \mathcal{N} (0, \Sigma_{f_1, \dots, f_m} )
     \,  ,
\end{align}
vaguely in distribution
, 
where $\mathcal{N} (0,\sigma_{f_1, \dots, f_m})$ denotes the centered  Gaussian with covariance matrix 
$\Sigma = (\Sigma_{i,j})_{i,j}$ given by 
\begin{align}
\label{eq:covariance}
\Sigma_{i,j} = \begin{cases}
\langle \sigma_{f_i}, \sigma_{f_j} \rangle & \text{if} \quad i<j \, , \\
\langle \sigma_{f_j}, \sigma_{f_i} \rangle & \text{otherwise} \, , 
\end{cases}
\end{align}
where the square integrable functions $\sigma_{f_i}$ are given by 
\begin{align}
\label{def:sigma-f}
\sigma_{f_i} (p) = \cosh (\mu_p) \widehat{(qf_i (O) \varphi)} (p) + \sinh (\mu_p) \widehat{(qf_i (O) \varphi)} (-p)
\end{align}
with $\varphi = 1_{\Lambda}$, $q = 1- \vert \varphi \rangle \langle \varphi \vert$ and for $\a0$ given by \eqref{def:a0}
\begin{align}
\label{def:mu}
\mu_p = \frac{1}{4} \log \left( \frac{p^2}{p^2 + 16 \pi \a0} \right) \quad \text{for all} \quad p \in \Lambda_+^* \, .
\end{align}
If in addition we assume that $f_1 , \dots, f_m \in \emph{Lip}_b(\R)$, then the convergence in \eqref{eq:CLT_main} holds weakly in distributions.
\end{enumerate}
\end{theorem}

The validity of a central limit theorem for the empirical measure $\mu_N$ given by Theorem \ref{thm:clt} provides a novel characterization of the particles outside the condensate wave function, often referred to as fluctuations around the condensate. The proof of this Theorem relies on a detailed description of these fluctuations discussed in more detail in Section \ref{sec:manybody}. \\

\paragraph{\textit{Lack of convergence of the infinite dimensional fields}}
The central limit theorem provided in Theorem~\ref{thm:clt} concerns the convergence of \textit{finite dimensional distributions} for the fluctuation field. 
A stronger result as the one provided by Theorem~\ref{thm:clt} would be to show that the centered, rescaled random measures
\begin{align}
    \sigma_N := \sqrt{N} 
    \big(
        \nu_N - \nu_\varphi
    \big) 
        : (\Omega, \mathbb P) \to \mathcal M_0(\R) 
\end{align} 
converge vaguely in distribution, with respect to some topology on the space of signed measures with zero mass, here denoted by $\mathcal M_0(\R)$. We want to stress that, when endowing such space with the total variation norm, a result of this type has no hope to hold, at least not for a general choice of $O$. Indeed, it is clear by construction that, if $O$ has a continuous spectrum, then the associated deterministic measure $\nu_\varphi$ might be singular with respect to the counting measure (e.g. if it is absolutely contiunous with respect to the Lebesgue measure), which makes it mutually singular with repsect to the empirical measure $\nu_N$, for every $N \in \N$. As a consequence, we would have that 
\begin{align}
    \left\|
        \sigma_N(\omega) 
    \right\|_{\text{TV}(\R)}
        = 
    2 \sqrt{N} \to \infty 
        \, , 
\end{align} 
for every $\omega \in \Omega$. Due to the fact that vaguely converging sequences of measures are necessarily bounded in total variation, it is not hard to show that this prevents any type of convergence in distribution for $\sigma_N$ with respect to the chosen topology. 
\\

\paragraph{\textit{Central limit theorems for $(Y_i^N)_{i=1}^N$}}  Note that fluctuations around the random variables $(Y_i^N)_{i=1}^N$ are known \cite{RS} to satisfy a central limit theorem with variance $\sigma^2 = \| \sigma_{\mathds{1}} \|_2^2$ where $\sigma_{\mathds{1}}$ is given by \eqref{def:sigma-f}. Recently \cite{R-LD2023}, the random variables $(Y_i^N)_{i=1}^N$ were further characterized through large deviation estimates in the regime of weak mean-field interactions, that is replacing the Gross--Pitaevskii interaction potential $V_N$ by 
\begin{align}
\label{eq:V-beta}
V_N^{(\beta)} = N^{3\beta} V (N^\beta \cdot)
\end{align}
for $\beta =0$ (i.e. $V_N^{(0)} = V$ independent of $N$) in the definition of $H_N$ in \eqref{eq:ham}. Note that the mean-field regime corresponds to weak, long-range particles interaction and is mathematically easier accessible then the physical most relevant Gross--Pitaevskii regime. The latter corresponds to $\beta =1$ and strong, singular particles' interaction. The so-called intermediate regime $\beta \in (0,1)$ interpolates between mean-field and Gross--Pitaevskii regime.  \\

\paragraph{\textit{Dynamics}} The present approach to Bose-Einstein condensates through random variables in the spirit of \eqref{eq:law-of-y} were first introduced for the dynamics of initial Bose-Einstein condensates. More precisely \cite{BKS} defines random variables $(Y_i^{N,t})_{i=1}^N$ with law given by \eqref{eq:law-of-y}, replacing the ground state $\psi_N$ by the solution $\psi_{N,t}$ of the Schr\"odinger equation 
\begin{align}
i \partial_t \psi_{N,t} = H_N \psi_{N,t}
\end{align}
with initial states $\psi_{N,0} = \varphi_0^{\otimes N}$. The random variables $(Y_i^{N,t})_{i=1}^N$ were proven to satisfy central limit theorems first for the mean-field, i.e. $\beta =0$ in \eqref{eq:V-beta}, \cite{BKS,BuSS,R}, later the intermediate ($\beta \in (0,1)$) \cite{R-2020} and recently for the Gross--Pitaevskii scaling regime ($\beta =1$) \cite{COS} . Large deviation estimates for $(Y_i^{N,t})_{i=1}^N$ however are only available for mean-field interactions ($\beta=0$) \cite{KRS,RS-JSP-22}. \\

\paragraph{\textit{Idea of the proof}} The proofs of Theorem \ref{thm:lln} and Theorem \ref{thm:clt} rely on a special feature of the $1$-Wasserstein distance that allows to write $W_1 ( \nu_N, \nu_\varphi)$ in terms of the respective cumulative distribution function (see beginning of Section \ref{susec:W1} for more details). With this observation, we show that the proofs of the theorems reduce to compute the leading order of the observable $\frac{1}{N} \sum_{i=1}^N g(O^{(i)}) - \langle \varphi,g( O) \varphi \rangle $ for any bounded $g$ in the ground state $\psi_N$ of $H_N$ (see Lemma \ref{lemma:variance}).  \\

\paragraph{\textit{Structure of the paper}} The rest of the paper is structured as follows: In Section \ref{sec:proof-thms} we prove Theorems \ref{thm:lln}, \ref{thm:clt}. To this end, we first explain in more detail the special feature of the $1$-Wasserstein distance (Section \ref{susec:W1}) that we then use to prove Theorem \ref{thm:lln}, Theorem \ref{thm:clt}(i) from Lemma \ref{lemma:variance} (in Section \ref{sec:proof-lln,clt1}) and later Theorem \ref{thm:clt}(ii) (in Section \ref{sec:clt2}). The rest of the paper is dedicated to the proof of Lemma \ref{lemma:variance}. For this we first give a brief introduction into the quantum many-body description of fluctuations around the condensate in Section \ref{sec:manybody} and then prove Lemma \ref{lemma:variance} in Section \ref{sec:lemma_variance}.

\section{Proofs of Theorems}
\label{sec:proof-thms}

\subsection{Preliminaries}

In this section we focus first on special properties of the $1$-Wasserstein distance and second on properties of the many-body ground state $\psi_N$ that we combine later in Section \ref{sec:proof-lln,clt1} to prove Theorems \ref{thm:lln},\ref{thm:lln} (i).  \\

 \label{susec:W1}
 \paragraph{\textit{Properties of the $1$-Wasserstein distance}} 
Here we focus on the enjoyable properties of Wasserstein distances when the underlying space is the real line as this is the setting relevant for our purposes. Given a Borel probability measure $\mu$ on the real line, its cumulative distribution function $F_{\mu}: \, \R \rightarrow [0,1]$ is given by $F_{\mu}(x)=\mu ((-\infty, x]).$ The \emph{quantile function} or monotone rearrangement of $\mu$ is denoted by $F_\mu^{-1}$ and is given by
\begin{equation} \label{eq:quantile-def}
F_{\mu}^{-1}: \, (0,1) \rightarrow \R; \, t \mapsto F_{\mu}^{-1}(t):=\sup\{x \in \R \, : \, F_\mu(x) \leq t \}.
\end{equation}
It is a general feature of the $p$-Wasserstein distances between probability measures on the real line that they are precisely the $L^p((0,1))$-distances between the corresponding quantile functions:
$$
W_p(\mu, \nu)=\norm{F_{\mu}^{-1}-F_{\nu}^{-1}}_{L^p((0,1))}=\left(\int_0^1 \abs{F_{\mu}^{-1}(t)-F_{\nu}^{-1}(t)}^p \dd t \right)^{\frac{1}{p}}.
$$
Furthermore, it is a special feature of the $1$-Wasserstein distances that they coincide with the $L^1(\R)$-distances of the respective cumulative distribution functions:
\begin{equation} \label{eq:W-1-dist}
W_1(\mu, \nu)=\norm{F_{\mu}-F_{\nu}}_{L^1(\R)}=\int_{\R}\abs{F_{\mu}(x)-F_{\nu}(x)} \dd x.
\end{equation}
The identity \eqref{eq:W-1-dist} will play an important role in our proofs of Theorem \ref{thm:lln} and Theorem \ref{thm:clt} (i) from the following Lemma.  \\

 \paragraph{\textit{Properties of the ground state $\psi_N$ of $H_N$}} The following Lemma, that is proven in Section \ref{sec:lemma_variance}, shows important properties of the ground state $\psi_N$. 

\begin{lemma}
\label{lemma:variance}
Let $V \in L^3 ( \R^3)$ be compactly supported, non-negative and spherically symmetric and $\psi_N$ denote the (uniquely) defined ground state of $H_N$ given by \eqref{eq:ham}. For any bounded $g\in L^\infty (\mathbb{R})$, there exists $C>0$ such that
\begin{align}
\label{eq:variance}
\Big\vert \langle \psi_N, \, \Big[ \frac{1}{N} \sum_{i=1}^N g(O^{(i)}) - \langle \varphi,g( O) \varphi \rangle \Big]^2 \psi_N \rangle - \frac{ \| \sigma_{g} \|_{\ell^2( \Lambda_+^*)^2}}{N} \Big\vert \leq \frac{C}{N^{5/4}}
\end{align}
where $C$ depends on $\| g \|_\infty$ and $\sigma_g$ is defined by \eqref{def:sigma-f}.
\end{lemma}

\subsection{Proof of Theorem \ref{thm:lln}, Theorem \ref{thm:clt} (i)} 
\label{sec:proof-lln,clt1} 
We use a slight abuse of notation and for the sake of simplicity write $F_\varphi := F_{\nu_\varphi}$. 
By Markov's inequality, Theorem \ref{thm:lln} is a consequence of Theorem \ref{thm:clt} (i) and thus it suffices to prove the latter one. 
% {
% \color{blue}
% Indeed, by Markov's inequality, one has  
% }
For this note that
\begin{align}
\bE \Big[W_1 ( \nu_N, \nu_\varphi ) \Big] =  \bE  \int_\mathbb{R} \vert F_{\nu_N} (x) - F_\varphi (x) \vert \dd x  = \bE  \int_{{\rm conv} ( \sigma (O))} \vert F_{\nu_N} (x) - F_\varphi (x) \vert \dd x \, ,
\end{align}
where at last we used that the fact that the support of $\nu_N$, $\nu_\varphi$ is a subset of $\rm conv(\sigma(O))$, which in particular implies that $ F_{\nu_N} (x) = F_\varphi (x)$, for every $x \notin \rm conv(\sigma(O))$, $N \in \N$. Using Fubini's theorem, we end up with
\begin{align}
\bE \Big[W_1 ( \nu_N, \nu_\varphi ) \Big] =  \int_{{\rm conv} ( \sigma (O))} \bE   \vert F_{\nu_N} (x) - F_\varphi (x) \vert \dd x \, , 
\end{align}
and by Cauchy-Schwarz inequality
\begin{align}
\bE \Big[W_1 ( \nu_N, \nu_\varphi ) \Big] \leq \left( \int_{{\rm conv} ( \sigma (O))} \dd x \right)^{1/2} \left( \int_{{\rm conv} ( \sigma (O))} \left( \bE \vert F_{\nu_N} (x) - F_\varphi (x) \vert \right)^2  \dd x  \right)^{1/2} \, .
\end{align}
Note that $\sigma (O)$ is a compact set on $\mathbb{R}$, thus ${\rm diam} ( \sigma (O) )< \infty$. Additionally, by Jensen's inequality,
\begin{align}
\bE \Big[W_1 ( \nu_N, \nu_\varphi ) \Big] \leq {\rm diam} ( \sigma (O))^{1/2} \left( \int_{{\rm conv} ( \sigma (O))}  \bE \vert F_{\nu_N} (x) - F_\varphi (x) \vert^2  \dd x  \right)^{1/2} \, .  \label{eq:W1-1}
\end{align} 
Secondly, we write explicitly the cumulative functions, i.e. using the special structure of $\nu_N$ and $\nu_\varphi$, together with the definition \eqref{eq:quantile-def}, we see that, for every $x \in \R$, $N \in \N$,  
\begin{align}
    F_{\nu_N}(x) 
        = 
    \frac{1}{N} \sum_{i=1}^N \mathds{1}_{Y^{(i)}\leq x } 
        \qquad \text{and} \qquad 
    F_\varphi(x) 
        =
    \langle \varphi, \mathds{1}_{(-\infty, x]} (O) \varphi \rangle
        \, . 
\end{align}
Plugging in these explicit expressions, setting by  $g_x := \mathds{1}_{(-\infty, x]}(\cdot )$,  we obtain that 
\begin{align}
    \bE  
    \vert 
        F_{\nu_N}  (x) - F_\varphi (x) 
    \vert^2    
        =
    \bE 
    \Big\vert 
        \frac{1}{N} \sum_{i=1}^N  \mathds{1}_{Y^{(i)}\leq x }
    \Big\vert^2  
        - 
      \bE 
    \Big[ 
        \frac{2}{N} 
        \sum_{i=1}^N \mathds{1}_{Y^{(i)} \leq x}
    \Big]  
    \langle 
        \varphi, g_x(O) \varphi 
    \rangle  
        +  
    \langle 
        \varphi, g_x(O) \varphi 
    \rangle^2
        \, .
\end{align}
Now, observe that by definition of the random variables $Y^{(i)}$ and their joint law, for every continuous map $f: \R^N \to \R$, the equality 
\begin{align}
    \bE f \big (Y^{(1)}, \dots, Y^{(N)} \big)
        =
    \langle
        \psi_N 
            , 
        f \big( O^{(1)}, \dots , O^{(N)} \big)
        \psi_N
    \rangle 
\end{align}
holds.
For clarity, we describe what we mean here and in the sequel by the action of a $k$-variable function on a $k$-tuple of commuting operators. Let $B_1, \dots, B_k$ be \emph{commuting} operators on a Hilbert space $\mathcal{H}.$ Then they admit a common spectral resolution, that is, a projection valued measure $\tilde{E}: \, \mathcal{B}(\R) \rightarrow \cP(\mathcal{H})$ such that $B_j=\int_{\R}\lambda_j(t) \dd \tilde{E}(t)$ for all $j \in \{1,2, \dots k\},$ where $\lambda_1, \dots, \lambda_k$ are measurable real functions. Now the action of $f: \R^k \rightarrow \R$ on $B_1, \dots, B_k$ is defined by 
$$
f(B_1, \dots, B_k)=\int_{\R} f\ler{\lambda_1(t), \lambda_2(t), \dots,\lambda_k(t)} \dd \tilde{E}(t).
$$

Therefore, we can proceed our estimates and write 
\begin{align}
    \bE \vert F_{\nu_N}   (x) - F_\varphi (x) \vert^2   
    &=
    \langle \psi_N, \,  \Big[ \frac{1}{N} \sum_{i=1}^N g_x(O^{(i)})\Big]^2  \psi_N \rangle  \\
     - 2  \langle \psi_N,  \frac{1}{N} &\sum_{i=1}^N g_x(O^{(i)}) \psi_N \rangle  \, \langle \varphi, g_x(O) \varphi \rangle
    +  \langle \varphi, g_x (O) \varphi \rangle_{L^2} \notag \\
    &= \langle \psi_N, \,  \Big[ \frac{1}{N} \sum_{i=1}^N g_x(O^{(i)})-\langle \varphi, g_x (O)\varphi \rangle_{L^2} \Big]^2  \psi_N \rangle \notag \\
    &= \langle \psi_N, \,  \Big[ \frac{1}{N} \sum_{i=1}^N g_x(O^{(i)})-\langle \varphi, g_x(O)\varphi \rangle\Big]^2  \psi_N \rangle
        \, . 
\end{align}
Note that $\| g_x \|_{L^\infty} \leq 1$ for all $x \in \mathbb{R}$ and thus, it follows from Lemma \ref{lemma:variance} that
\begin{align}
\Big\vert \bE \vert F_{\nu_N} (x) - F_\varphi (x) \vert^2 - \frac{\| \sigma \|_{\ell^2( \Lambda_+^*)}^2}{N} \Big\vert \leq \frac{C}{N^{5/4}} \, .
\end{align}
Plugging this into \eqref{eq:W1-1}, we arrive at Theorem \ref{thm:clt}(i).
\qed

\subsection{Proof of Theorem \ref{thm:clt} (ii)}
\label{sec:clt2}
Let $f_1, \dots, f_m$ be continuous and bounded functions. In order to show the sought convergence in distribution (vaguely), we prove that for all $ G \in C_c ( \mathbb{R}^m)$ we have
\begin{align}
    \lim_{N \rightarrow \infty} 
    \bE 
    \left[ 
        G  
        \left\{ 
            \sqrt{N} 
            \big( 
                \langle \nu_N (\omega),  f_j \rangle - \langle \nu_\varphi,  f_j \rangle 
            \big) 
        \right\}_{j=1}^m \right] =  \int G ( \lambda ) \, \varrho_{0, \Sigma_{f_1, \dots, f_m}} (\lambda ) d \lambda
\end{align}
where $\varrho_{0, \Sigma_{f_1, \dots, f_m}}$ denotes the centered Gaussian with covariance matrix $\Sigma =\Sigma(f_1, \dots, f_m)$ as given in \eqref{eq:covariance}. Note that by definition, we can write the l.h.s. as
\begin{align}
% \bE \left[ G  \lbrace \sqrt{N} \left( \langle \nu_N (\omega),  f_j \rangle - \langle \nu_\varphi,  f_j \rangle \right) \rbrace_{j=1}^m \right]  
% =
    \Big\langle 
        \psi_N,  G
        \Big( 
            \Big\{
                \sqrt{N}
                \Big(
                    \frac{1}{N}\sum_{i=1}^N f_j(O^{(i)}) - \langle \varphi, f_j( O) \varphi \rangle 
                \Big) 
            \Big\}_{j=1}^m 
        \Big) 
        \psi_N 
    \Big\rangle  \, .
\end{align}
% {\color{red} 
% Since $O$ is a bounded operator, the spectrum $\sigma (O)$ is a compact set and thus, it suffices to consider $G \in C_b( \mathbb{R}^m)$ with compact support (on $\sigma (O)$). [LORE: Is this true? Does not look right to me]
% }
Denoting by $\hat G$ the Fourier transform of $G \in L^1( \mathbb{R}^m)$, we can then write
\begin{align}
    \bE &
    \left[ 
        G  
        \left\{ 
            \sqrt{N} 
            \big( 
                \langle \nu_N (\omega),  f_j \rangle - \langle \nu_\varphi,  f_j \rangle 
            \big) 
        \right\}_{j=1}^m \right] \\
    &= \int d\lambda_1 \dots d\lambda_m \, \widehat{G}( \lbrace \lambda_j \rbrace_{j=1}^m ) \, \langle \psi_N,  e^{ i \sum_{j=1}^m  \lambda_j  \sqrt{N}(\frac{1}{N}\sum_{i=1}^N f_j(O^{(i)}) - \langle \varphi, f_j( O) \varphi \rangle ) }\psi_N \rangle  \, \notag \\
    &= \int d\lambda_1 \dots d\lambda_m \, \widehat{G}( \lbrace \lambda_j \rbrace_{j=1}^m ) \, \langle \psi_N,  \prod_{j=1}^m e^{ i \lambda_j  \sqrt{N}(\frac{1}{N}\sum_{i=1}^N f_j(O^{(i)}) - \langle \varphi, f_j( O) \varphi \rangle ) }\psi_N \rangle
        \, ,
\end{align}
where we used that the operators $f_j (O^{(i)})$ commute. We split the proof of the convergence into two steps, first making an extra assumption on $\hat G$ and then removing this assumption by means of an approximation argument. 

\smallskip 
\noindent 
\textit{Step 1}. \ 
We first make the extra assumption that $G \in L^1(\R^m)$ also satisfies 
\begin{align}
\label{eq:condition_Ghat}
    \hat G \in L^1 
    \big( 
        \mathbb{R}^m, (1+ \vert s_1 \vert^4 + \dots \vert s_m \vert^4) ds_1 \dots ds_m  
    \big)
        \,  , 
\end{align} 
which essentially corresponds to a regularity assumption on $G$, see Step 2. Arguing as in the proof of \cite[Theorem~1.1]{RS}, one can show that for any such $G$ the estimate
\begin{align}
\big \vert   \bE & \left[ G  \lbrace \sqrt{N} \left( \langle \nu_N (\omega),  f_j \rangle - \langle \nu_\varphi,  f_j \rangle \right) \rbrace_{j=1}^m \right] - \int_{\mathbb{R}^m} \varrho_{0, \Sigma_{f_1, \dots f_m}} ( \lambda_1, \dots, \lambda_m )\prod_{j=1}^k g( \lambda_j) d\lambda_j \big\vert \notag \\
\leq&  CN^{-1/4}  \prod_{j=1}^m \int  \vert \widehat{G}_j (s_1, \dots s_m) \vert \,  (1+ \vert s_1 \vert^4 + \dots \vert s_m \vert^4) ds_1 \dots ds_m  \, .
\end{align}
holds true, for some constant $C$ which is independent of $N$. Sending $N \to \infty$, we conclude the sough convergence for $G$ belonging to a subclass of $C_c(\R^m)$, namely the one satisfying the additional condition \eqref{eq:condition_Ghat}. 

\smallskip 
\noindent 
\textit{Step 2}. \ 
We conclude the proof by approximation. Indeed, we observe that
\begin{align*}
    C_c^4(\R^m) \subset
    \left\{
        G \in L^1(\R^m)
            \suchthat
        \hat G \in L^1( \mathbb{R}^m, (1+ \vert s_1 \vert^4 + \dots \vert s_m \vert^4) ds_1 \dots ds_m  )
    \right\}
\end{align*}
and that $C_c^4(\R^m)$ is dense in $C_c(\R^m)$ with respect to the uniform convergence. Due to the fact that $\{ \law(X_N) \}_N \subset \cP(\R^m)$ (hence they have uniformly bounded total mass), a classical argument shows that the sought convergence holds for all $G \in C_c(\R^m)$. Indeed, for $G \in C_c(\R^m)$, let $G_k \to G$ uniformly as  $k \to \infty$, with $G_k \in C_c^4(\R^d)$. Then we have that
\begin{align*}
    \left|
        \bE[G(X_N)]
            -
        \bE[G_k(X_N)]
    \right|
        \leq
    \norm{G-G_k}_\infty
        \, ,
\end{align*}
for every $k,N$. Now, when we take the limit as $N \to \infty$ for a fixed $k$, we already know that $ \bE[G_k(X_N)] \to  \bE[G_k(X_\infty)]$ as $N \to \infty$, where $X_\infty$ is a Gaussian random variable with the claimed covariance structure. Therefore, from the previous estimate we have, for every $k \in \N$,
\begin{align*}
    \limsup_{N \to \infty}
    &\left|
        \bE[G(X_N)]
            -
        \bE[G(X_\infty)]
    \right|
\\
        &\qquad \leq
 \limsup_{N \to \infty}
    \left|
        \bE[G(X_N)]
            -
        \bE[G_k(X_N)]
    \right|
    +
     \left|
        \bE[G_k(X_N)]
            -
        \bE[G_k(X_\infty)]
    \right|
\\
    &\qquad \qquad \qquad \qquad 
        +
    \left|
        \bE[G_k(X_{\infty})]
            -
        \bE[G(X_\infty)]
    \right|
\\
        &\qquad \leq
    \norm{G-G_k}_\infty
        +
    0
    +\norm{G-G_k}_\infty
        =
    2 \norm{G-G_k}_\infty
        \, .
\end{align*}
Sending $k \to \infty$, we find the sought weak convergence
\begin{align*}
    \lim_{N \to \infty}
         \bE[G(X_N)]
            =
        \bE[G(X_\infty)]
            \, ,
\end{align*}
thus concluding the proof.
% \textcolor{magenta}{How to conclude here? Somehow we have to find a proper appoximation of $G$ (maybe in the spirit of \cite[Section 3]{R}? There is proven an approximation of $\mathds{1}_[a,b]$ of an interval $[a,b]$ by sufficiently nice functions) }

\smallskip 
\noindent 
\textit{Tightness}. \ 
We are left with the proof of the convergence weakly in distribution in the case when $f_1, \dots, f_m$ are Lipschitz and bounded. From the previous part of the proof, we already know that the corresponding laws converge vaguely to the sought Gaussian limit. Therefore, by Prokohorov's theorem, it is enough to show that the law of the random vector is tight, under the assumption of Lipschitzianity. More precisely, we claim that the laws of the random vector $ X_N:= \lbrace \sqrt{N} \left( \langle \nu_N (\omega),  f_j \rangle - \langle \nu_\varphi,  f_j \rangle \right) \rbrace_{j=1}^m \in \mathbb R^m$ as probability measures in $\mathbb R^m$ is tight, and therefore weakly (pre)compact. Denote by $B_R$ the closed ball centered in $0$ in $\mathbb R^m$ of radius $R>0$. Recall the Kantorovich duality formula for $\mathbb W_1$, which states that 
\begin{align}
    \mathbb{W}_1(\mu_0,\mu_1) 
        =
    \sup
    \left\{
        \frac1{\text{Lip}(\varphi)}\int \varphi \dd (\mu_1 - \mu_ 0) 
            \suchthat 
        \varphi \in \text{Lip}(\R) \cap C_b(\R) 
    \right\}
        \, .
\end{align}
As a consequence of duality and thanks to Markov's inequality, we conclude that
\begin{align*}
	\text{Law}(X_N) (B_R^c) = \mathbb{P}
	\big(
		\| X_N \| > R
	\big)
		&\leq 	
	\frac1R
	\bE
	\big(
		\| X_N \|
	\big)
		\leq 	
	\frac1R \sqrt{N}
	\bE
	\big(
		\sup_{j = 1, \dots, m}
		\big|
			 \langle \nu_N (\omega),  f_j \rangle - \langle \nu_\varphi,  f_j \rangle
		\big|
	\big)
\\	
		&\leq
	\big( \sup_{j=1, \dots, m} \text{Lip}(f_j) \big)
	\frac1R \sqrt{N}
	\bE \left[ W_1 ( \nu_N, \nu_\varphi) \right]
		\leq \frac CR
	\, ,
\end{align*}
for every $f_1, \dots, \dots f_m$ Lipschitz and bounded, where at last we used (i). The closed ball in $\mathbb R^m$ being compact, this provides the claim tightness (by taking $R \to \infty$).

\section{Fluctuations around the condensate}
\label{sec:manybody}

To prove Lemma \ref{lemma:variance}, we need a quantitative estimate for the number of partciles outside the condensate, called excitations. For this we factor out the condensate's contributions using that any $\psi \in L^2( \Lambda^N)$ can be uniquely written as
\begin{align}
\psi = \eta \varphi^{\otimes N} + \eta_1 \otimes_s \varphi^{\otimes (N-1)} + \dots +\eta_N
\end{align}
with $\eta_j \in L_{\perp \varphi}^2 ( \Lambda)^{\otimes_s}$, where $\otimes_s$ denotes the symmetrized tensor product and 
$L_{\perp \varphi}^2 ( \Lambda)$ 
the orthogonal complement in $L^2( \Lambda)$ of the condensate wave function. This observation allows to define the unitary map 
\begin{align}
\label{def:U}
\mathcal{U}_N: L_s^2( \Lambda^N) \rightarrow \mathcal{F}_{\perp \varphi}^{\leq N} := \bigoplus_{n=0}^N L^2_{\perp \varphi} ( \Lambda)^{\otimes_s n}, \, \quad \psi \mapsto \lbrace \eta_1, \dots, \eta_N \rbrace \, .
\end{align}
In momentum space $\Lambda^* = (2\pi) \mathbb{Z}^3$, the condensate wave function corresponds to zero momentum $p=0$, thus the excitation vector $ \mathcal{U}_N\xi_N \in \mathcal{F}_{\perp \varphi}^{\leq N}$ describes particles with momenta in $\Lambda_+^* := \Lambda^*\setminus \lbrace 0 \rbrace$. The truncated Fock space $\mathcal{F}_{\perp \varphi}^{\leq N}$ is equipped with standard creation and annihilation operators $a^*_p, a_q$ with $p,q \in \Lambda_+^*$  satisfying the canonical commutation relations
\begin{align}
\label{eq:CCR}
\left[ a^*_p, a^*_q \right] = \left[ a_p, a_q \right]  = 0, \quad \text{and} \quad \left[ a^*_p, a_q \right]  = \delta_{p,q} \, .
\end{align}
The unitary $\mathcal{U}_N$ defined in \eqref{def:U} satisfies for $p,q \in \Lambda_+^*$ the properties \cite{LNSS}
\begin{align}
\label{eq:prop-U}
\mathcal{U}_N^* a^*_0 a_0 \mathcal{U}_N &=  N - \mathcal{N}_+ \notag 
    \, ,
\\
\mathcal{U}_N^* a^*_p a_q \mathcal{U}_N &= a^*(f) a(g)
    \, , 
\\
\mathcal{U}_N^* a^*_p a_0 \mathcal{U}_N &= a^*_p \sqrt{N - \mathcal{N}_+} =: \sqrt{N} b^*_p \notag
    \, , 
\\
\mathcal{U}_N^* a^*_0 a_p \mathcal{U}_N &=  \sqrt{N - \mathcal{N}_+} a_p =: \sqrt{N} b_p  \, , 
\end{align}
where the number of excitations $\mathcal{N}_+$ is given by
\begin{align}
\mathcal{N}_+ := \sum_{p \in \Lambda_+^*} a_p^*a_p \, .
\end{align}
Moreover we introduced  for $p,q \in \Lambda_+^*$ the modified creation and annihilation operators $b^*_p, b_q$ that, in contrast to the standard creation and annihilation operators $a^*_p, a_q$, leave the truncated Fock space $\mathcal{F}_{\perp \varphi}^{\leq N}$ invariant (as $b^*_p \psi =0$  for any $\psi \in L_{\perp \varphi} ( \Lambda_+^*)^N$). For states with a bounded number of excitations $\mathcal{N}_+$, the modified creation and annihilation operators $b_p^*,b_p$ asymptotically behave as standard one $a_p^*,a_p$ when $N \rightarrow \infty$. Their commutation relations
\begin{align}
\label{eq:mCCR}
\left[ b^*_p, b^*_q \right] = \left[ b_p, b_q \right]  = 0, \quad \text{and} \quad \left[ b^*_p, b_q \right]  = \delta_{p,q} \Big(1-  \frac{\cN_+}{N} \Big) - \frac{1}{N} a_p^*a_q
\end{align}
agree with the standard canonical commutation relations \eqref{eq:CCR} up to an error of order $N^{-1}$. Denoting for $h \in \ell^2( \Lambda_+^*)$
\begin{align}
b(h) :=\sum_{p \in \Lambda_+^*} \overline{h}_p b_p , \quad b^*(h) := \sum_{p \in \Lambda_+^*} h_p b_p^* \, , 
\end{align}
we then have  \cite[p. 235]{BBCS}
\begin{align}
\label{eq:bounds.b}
\| b(h) \psi \| \leq \| h \|_{\ell^2 ( \Lambda_+^*)} \| \mathcal{N}_+^{1/2} \psi \|, \quad  \| b^*(h) \psi \| \leq \| h \|_{\ell^2( \Lambda_+^*)} \| (\mathcal{N}_+ + 1)^{1/2} \psi \| \, .
\end{align}
For the ground state, the number of excitations w.r.t. to the condensate (that is the number of of the excitation vector $\xi_N = \mathcal{U}_N \psi_N$) is of order one; more precisely, \cite[Lemma 2.1 and Proposition 4.1]{BBCS} show that for $k \in \mathbb{N}$
\begin{align}
\label{eq:bound-N}
\langle \mathcal{U}_N \psi_N, \,  ( \mathcal{N}_+ + 1)^k \mathcal{U}_N \psi_N \rangle  \leq C_k
\end{align}
and some constant $C_k>0$. Furthermore, \cite{BBCS} provides an approximation of the excitation vector $\xi_N \in \mathcal{F}_{\perp \varphi}^{\leq N}$ in $L^2( \Lambda)^N$-norm. For this, the particles' correlation, that is inherited in the excitation vector, plays a crucial role. The correlation structure is described mathematically through a generalized Bogoliubov transform; a unitary map $e^B( \eta) : \mathcal{F}_{\perp \varphi}^{\leq N} \rightarrow \mathcal{F}_{\perp \varphi}^{\leq N}$ given by
\begin{align}
e^{B( \eta)} = \exp \Big( \frac{1}{2} \sum_{p \in \Lambda_+^*} \eta_p \left[ b_p^*b_{-p}^* - b_{p}b_{-p} \right]  \Big)
    \, , 
\end{align}
with $\eta_\cdot \in \ell^2 ( \Lambda_+^*) $ (see \cite[Lemma 3.1]{BBCS}) defined by
\begin{align}
\eta_p = - \frac{1}{N^2} \widehat{w}_\ell (p/N) , \quad \text{with} \quad \widehat{w}_\ell (k) = \int_{\mathbb{R}^3} \dd x \,  w(x) e^{ik x}
\end{align}
and $w_\ell = 1- f_\ell$, is given in terms of the solution $f_\ell$ of the Neumann problem
\begin{align}
\Big[- \Delta + \frac{1}{2} V \Big] f_\ell = \lambda_\ell f_\ell
\end{align}
on the ball $\vert x \vert \leq N \ell$ for an $\ell \in (0,\frac{1}{2})$. The generalized Bogoliubov transform acts onto modified creation and annihilation operators for $h \in \ell^2( \Lambda_+^*)$ as
\begin{align}
\label{eq:action-bogo}
e^{B( \eta)} b^*(h) e^{-B( \eta)} =& \sum_{p \in \Lambda_+^*}  \left[\cosh (\eta_p) h_p b_p^* + \sinh ( \eta_{p} ) h_p b_{-p} \right] + d^*_\eta (h), \notag \\
 e^{B( \eta)} b(h) e^{-B( \eta)} =& \sum_{p \in \Lambda_+^*}  \left[\cosh (\eta_p ) h_p b_p + \sinh (\eta_{p} ) h_p b_{-p}^* \right] + d_\eta (h)
    \, , 
\end{align}
where  $d_\eta (h), d^*_\eta (h)$ are small on states with a bounded number of excitations. From \cite[Lemma 2.3]{BBCS} it follows that for any $k \in \mathbb{Z}$ we have
\begin{align}
\label{eq:bound-d}
\| ( \mathcal{N}_+ + 1)^{k} d_\eta (h) \xi \| \leq  \frac{C_k}{N} \left(  \| h \|_{\ell^2( \Lambda_+^*)} + \| \eta \|_{\ell^2( \Lambda_+^*)} \right) \| ( \cN_+ + 1)^{3/2+k} \xi \| \, .
\end{align}
for some $C_k>0$. Furthermore, the generalized Bogoliubov transform leaves the number of excitation approximately invariant, i.e. for $k \in \mathbb{N}$ there exists $C_k >0$ such that
\begin{align}
\label{eq:bogo-N}
e^{B(\eta)} \cN_+^k e^{-B(\eta)} \leq C  (\cN_+ + 1)^k   \, .
\end{align}
The excitation Hamiltonian $\mathcal{G}_N := e^{-B(\eta)} \mathcal{U}_N H_N  \mathcal{U}_N^* e^{B( \eta)}$ can be decomposed \cite[Proposition 3.2]{BBCS} as $\mathcal{G}_N = C_N + \mathcal{Q}_N + \mathcal{C}_N + \mathcal{V}_N + \mathcal{E}_N$. Here, $C_N$ is a constant (and equals $4 \pi a_0 N$ up to errors that are order one in the limit $N \rightarrow \infty$), $\mathcal{Q}_N$ is quadratic, $\mathcal{C}_N$ cubic, and $\mathcal{V}_N$ quartic in (modified) creation and annihilation operators; the remainder $\mathcal{E}_N$ is small on low-energy states. In the Gross--Pitaevski regime the cubic term $\mathcal{C}_N$ (contrarily to other scaling regimes of less singular particles interaction) contributes to energy of order one. To extract the leading-order (order one) contribution of $\mathcal{C}_N$, the excitation Hamiltonian $\mathcal{G}_N$ is furthermore conjugated with a unitary cubic in modified creation and annihilation operators and given by
\begin{align}
e^{-A} = \exp \Big( \frac{1}{\sqrt{N}} \sum_{r \in P_H, v \in P_L } \eta_r \left[ \sinh( \eta_v) b^*_{r+v} b_{-r}^* b_{-v}^* + \cosh( \eta_v) b^*_{r+v} b_{-r}^* b_{-v}^* - {\rm h.c.} \right] \, , 
\end{align}
where h.c. denotes the hermitian conjugate of the previous operator and $P_L := \lbrace p \in \Lambda_+^*: \vert p \vert \leq N^{1/2} \rbrace$ describes low and $P_H = \Lambda_+^* \setminus \lbrace 0 \rbrace$ high momenta (by definition $r+v \not= 0$). Furthermore, similarly to \eqref{eq:bogo-N} it approximately preserves the number of particles, i.e. for any $k \in \mathbb{N}$ there exists $C>0$ such that
\begin{align}
\label{eq:eA-N}
e^{A} \cN_+^k e^{-A} \leq C  (\cN_+ + 1)^k  \, .
\end{align}
The conjugated excitation Hamiltonian $\mathcal{J}_N :=  e^{-A}e^{-B(\eta)} \mathcal{U}_N H_N  \mathcal{U}_N^* e^{B( \eta)} e^{A} $ has the decomposition  $\mathcal{M}_N = \widetilde{C}_N + \widetilde{\mathcal{Q}}_N + \mathcal{V}_N + \widetilde{\mathcal{E}}_N$ (see \cite[Proposition 3.3]{BBCS}) for a constant $\widetilde{C}_N$, a remainder $\widetilde{\mathcal{E}}_N$ small on low energy states, a quartic (positive) contribution $\mathcal{V}_N$ and a quadratic term given by
\begin{align}
\widetilde{\mathcal{Q}}_N  =\sum_{p \in \Lambda_+^*} F_p a_p^*a_p + \frac{1}{2} \sum_{p \in \Lambda_+^*} G_p \left[ b_p^*b_{-p}^* + b_p b_{-p} \right]
\end{align}
with
\begin{align}
F_p =& p^2 ( \sinh_{\eta_p}^2 + \cosh_{\eta_p}^2 ) + ( \widehat{V} (\cdot/N) * \widehat{f}_{N, \ell} )_p ( \sinh_{\eta_p} + \cosh_{\eta_p})^2, \notag \\
G_p =& p^2 \sinh_{\eta_p} \cosh_{\eta_p}  + ( \widehat{V} (\cdot/N) * \widehat{f}_{N, \ell} )_p ( \sinh_{\eta_p} + \cosh_{\eta_p})^2 \, .
\end{align}
Finally, to conclude with an approximation of the ground state energy, the excitation is conjugated with another (generalized) Bogoliubov transform $e^{B( \tau )}$ with
\begin{align}
\label{eq:tau}
\tanh ( 2 \tau_p) = - G_p / F_p
\end{align}
that (approximately) diagonalizes the quadratic term $\widetilde{\mathcal{Q}}_N$. Note that
\begin{align}
\label{eq:bounds-tau}
\| \tau \|_{\ell^2( \Lambda_+^*)} \leq C, \quad \| \tau \|_{\ell^\infty( \Lambda_+^*)} \leq C
\end{align}
for a constant $C>0$ independent of $N$ (see \cite[Lemma 5.1]{BBCS}). Thus, similarly to \eqref{eq:bogo-N} for $k \in \mathbb{N}$ there exists $C_k >0$ such that
\begin{align}
\label{eq:bogo-N-2}
e^{B(\tau)} \cN_+^k e^{-B(\tau)} \leq C  (\cN_+ + 1)^k   \, .
\end{align}
The final excitation Hamiltonian $\mathcal{J}_N = e^{-B ( \tau)} e^{-A}e^{-B(\eta)} \mathcal{U}_N H_N  \mathcal{U}_N^* e^{B( \eta)} e^{A} e^{B ( \tau)}$ has the form $\mathcal{J}_N = \widetilde{\widetilde{C}}_N + \mathcal{D}_N + \mathcal{V}_N + \widetilde{\widetilde{\mathcal{E}}}_N$, where $\mathcal{D}_N$ is quadratic and diagonal operator, $\mathcal{V}_N$ is quartic and positive and the remainder $\widetilde{\widetilde{\mathcal{E}}}_N$ is small on low-energy states. Thus the vacuum vector $\Omega = \lbrace 1, 0, \cdots \rbrace$ provides an approximation of the ground state of the excitation Hamiltonian $\mathcal{J}_N$. More precisely, \cite[Section 6]{BBCS} shows that there exists a phase $\omega \in [0, 2 \pi]$ such that
\begin{align}
\label{eq:norm-approx}
\| \mathcal{U}_N \psi_N  - e^{i \omega} e^{B( \eta)} e^{-A} e^{B(\tau)} \Omega \| \leq  C N^{-1/4} \, .
\end{align}

\section{Proof of Lemma \ref{lemma:variance}}
\label{sec:lemma_variance}

\begin{proof} To prove \eqref{eq:variance}, we first introduce the notation
\begin{align}
\label{def:tildeg}
\widetilde{g} (O^{(i)} ) :=  g(O^{(i)}) - \langle \varphi,g( O) \varphi \rangle
\end{align}
and observe that we can then write the l.h.s. of \eqref{eq:variance} as
\begin{align}
\label{eq:variance-1}
\langle \psi_N, \, \Big[ \frac{1}{N} \sum_{i=1}^N g(O^{(i)}) - \langle \varphi,g( O) \varphi \rangle \Big]^2 \psi_N \rangle = \langle \psi_N, \, \Big[ \frac{1}{N} \sum_{i=1}^N \widetilde{g} (O^{(i)} ) \Big]^2 \psi_N \rangle \, .
\end{align}
Furthermore we can express the above expression in terms of the second quantization of the operator $\widetilde{g} (O ) $: the second quantization of $\widetilde{g} (O ) $ is the operator $d \Gamma (\widetilde{g} (O ))$ on $ \mathcal{F} := \bigoplus_{n \geq 0}^{\infty} L^2 ( \Lambda)^{\otimes_s n}$ defined by the requirement that on the $n$-particle sector $\psi^{(n)} \in  L^2 ( \Lambda)^{\otimes_s n}$ of the Fock-space vector $ \psi = \bigoplus_{n \geq 0}^\infty \psi^{(n)}\in \mathcal{F}$, it holds
\begin{align}
\left( d \Gamma (\widetilde{g} (O )) \psi \right)^{(n)} = \sum_{i=1}^n  \widetilde{g} (O)^{(i)} \psi^{(n)} = \sum_{i=1}^n  \widetilde{g} (O^{(i)}) \psi^{(n)} \, .
\end{align}
Since $\psi_N \in L_s^2 (\Lambda)^N$ we thus have for \eqref{eq:variance-1}
\begin{align}
\langle \psi_N, \, \Big[ \frac{1}{N} \sum_{i=1}^N g(O^{(i)}) - \langle \varphi,g( O) \varphi \rangle \Big]^2 \psi_N \rangle =  \frac{1}{N^2}\langle \psi_N, \, \Big[ d \Gamma ( \widetilde{g} (O))\Big]^2 \psi_N \rangle \, .
\end{align}
Embedding $\psi_N \in L^2_s ( \Lambda)^N$ in the full bosonic Fock space $\mathcal{F}$, we use the identity
\begin{align}
d\Gamma (\widetilde{g} (O)) = \int \dd x\dd y \, ( \widetilde{g} (O) )(x;y)  \, a_x^*a_y
    \, , 
\end{align}
where $a_x^*,a_y$ denote the standard creation and annihilation operators and $\widetilde{g}(O)(x;y)$ denotes the kernel of the operator $\widetilde{g} (O)$. 
This representation of $d\Gamma (\widetilde{g} (O))$ is in fact useful to compute the action of the excitation map $\mathcal{U}_N$ on $d\Gamma (\widetilde{g} (O))$ in the next step. To this end, we write using that $\mathcal{U}_N$ is a unitary operator
\begin{align}
\label{eq:variance-2}
\langle \psi_N, \, \Big[ \frac{1}{N} \sum_{i=1}^N g(O^{(i)}) - \langle \varphi,g( O) \varphi \rangle \Big]^2 \psi_N \rangle =&  \frac{1}{N^2}\langle \mathcal{U}_N\psi_N, \mathcal{U}_N \, \Big[ d \Gamma ( \widetilde{g} (O))\Big]^2 \mathcal{U}_N^* ( \mathcal{U}_N \psi_N  ) \rangle \, \notag \quad \\
=&\frac{1}{N^2}\langle \mathcal{U}_N\psi_N,  \, \Big[ \mathcal{U}_N d \Gamma ( \widetilde{g} (O)) \mathcal{U}_N^* \Big]^2  ( \mathcal{U}_N \psi_N  ) \rangle \, .
\end{align}
The properties \eqref{eq:prop-U} of the unitary $\mathcal{U}_N$ allow to compute $\mathcal{U}_N d \Gamma ( \widetilde{g} (O)) \mathcal{U}_N^*$ explicitly. To this end, we first decompose $ \widetilde{g} (O) $ w.r.t. 
$\pi = \vert \varphi \rangle \langle \varphi \vert $ and find
\begin{align}
\widetilde{g} (O) =& \pi \widetilde{g} (O) \pi + \pi \widetilde{g} (O) (1- \pi) +  (1- \pi) \widetilde{g} (O) \pi +  (1- \pi) \widetilde{g} (O)  (1- \pi)\notag \\
 =&   \pi \widetilde{g} (O)  (1- \pi) +  (1- \pi) \widetilde{g} (O) \pi +  (1- \pi) \widetilde{g} (O)  (1- \pi)   \, ,
\end{align}
where the second line follows from the definition of $\widetilde{g} (O)$ in \eqref{def:tildeg}, that implies $\pi \widetilde{g} (O) \pi = 0$. With the notation
\begin{align}
h :=   (1- \pi)  \widetilde{g} (O)  \varphi, \quad H:=   (1- \pi) \widetilde{g} (O)  (1- \pi)
\end{align}
and \eqref{eq:prop-U} we thus find 
\begin{align}
\label{def:h,H}
\mathcal{U}_N d \Gamma ( \widetilde{g} (O)) \mathcal{U}_N^* = \sqrt{N} b(h) +  \sqrt{N}b^*(h) + d \Gamma (H) \, .
\end{align}
Plugging this into \eqref{eq:variance-2}, we get
\begin{align}
\langle \psi_N,  & \, \Big[ \frac{1}{N} \sum_{i=1}^N g(O^{(i)}) - \langle \varphi,g( O) \varphi \rangle \Big]^2 \psi_N \rangle \notag \\=& \frac{1}{N^2} \langle\mathcal{U}_N \psi_N, \Big[ \sqrt{N} b(h) + \sqrt{N} b^* (h) + d\Gamma (H) \Big]^2 \mathcal{U}_N \psi_N \rangle \notag \\
=& \frac{1}{N} \langle\mathcal{U}_N  \psi_N, \Big[ b^*(h) + b(h) \Big]^2 \mathcal{U}_N \psi_N \rangle \notag \\
&+ \frac{1}{N^{3/2}} \langle \mathcal{U}_N  \psi_N, \, \left( \Big[ b^*(h) + b(h) \Big] d \Gamma (H) + d \Gamma (H) \Big[ b^*(h) + b(h) \Big] \right) \mathcal{U}_N \psi_N \rangle  \notag \\
&+\frac{1}{N^2} \langle\mathcal{U}_N  \psi_N, \Big[d\Gamma (H) \Big]^2 \mathcal{U}_N \psi_N \rangle \label{eq:variance-3} \, .
\end{align}

\subsubsection*{Step 1 (Action of $\mathcal{U}_N$)} 
Next, we show that the last two terms are $O(N^{-3/2})$, i.e. sub-leading for our analysis, while we extract the leading order contribution from the first term. More precisely we claim that there exists $C>0$ such that
\begin{align}
\Big\vert \langle \psi_N,  & \, \Big[ \frac{1}{N} \sum_{i=1}^N g(O^{(i)}) - \langle \varphi,g( O) \varphi \rangle \Big]^2 \psi_N \rangle \notag -  \frac{1}{N} \langle\mathcal{U}_N  \psi_N, \Big[ b^*(h) + b(h) \Big]^2 \mathcal{U}_N \psi_N \rangle \Big\vert  \leq \frac{C \| g \|_\infty}{N^{3/2}} \, .
\end{align}
To this end, we observe that from \eqref{eq:variance-3} one infers
\begin{align}
\Big\vert \langle \psi_N,  & \, \Big[ \frac{1}{N} \sum_{i=1}^N g(O^{(i)}) - \langle \varphi,g( O) \varphi \rangle \Big]^2 \psi_N \rangle \notag -  \frac{1}{N} \langle\mathcal{U}_N  \psi_N, \Big[ b^*(h) + b(h) \Big]^2 \mathcal{U}_N \psi_N \rangle \Big\vert  \notag \\
&\leq \frac{1}{N^{3/2}} \langle \mathcal{U}_N  \psi_N, \, \left( \Big[ b^*(h) + b(h) \Big] d \Gamma (H) + d \Gamma (H) \Big[ b^*(h) + b(h) \Big] \right) \mathcal{U}_N \psi_N \rangle  \notag \\
&\quad +\frac{1}{N^2} \langle\mathcal{U}_N  \psi_N, \Big[d\Gamma (H) \Big]^2 \mathcal{U}_N \psi_N \rangle \, . \label{eq:variance-4}
\end{align}
We bound the two of the r.h.s. of \eqref{eq:variance-4} separately. For the first term, we write
\begin{align}
\frac{1}{N^{3/2}}  & \vert \langle \mathcal{U}_N  \psi_N, \, \left( \Big[ b^*(h) + b(h) \Big] d \Gamma (H) + d \Gamma (H) \Big[ b^*(h) + b(h) \Big] \right)\mathcal{U}_N  \psi_N \rangle\vert \notag \\
\leq& \frac{2}{N^{3/2}} \left( \| b^*(h) \mathcal{U}_N  \psi_N \| + \| b(h) \mathcal{U}_N  \psi_N \| \right)  \| d\Gamma ( H)  \mathcal{U}_N  \psi_N \| \, .
\end{align}
With \eqref{eq:bounds.b} and $\| d \Gamma ( J) \psi \| \leq \| J \| \| \mathcal{N} \psi \|$ 
for any bounded $J$ and $\psi \in \mathcal{F}$ we furthermore find
\begin{align}
\frac{1}{N^{3/2}}  & \vert \langle \mathcal{U}_N  \psi_N, \, \left( \Big[ b^*(h) + b(h) \Big] d \Gamma (H) + d \Gamma (H) \Big[ b^*(h) + b(h) \Big] \right) \mathcal{U}_N \psi_N \rangle\vert \notag \\
\leq&  \frac{4}{N^{3/2}} \| h \|_{L^2_{\perp \varphi} ( \Lambda )}  \| H \| \| ( \mathcal{N}_+ + 1)^{1/2} \mathcal{U}_N \psi_N \| \, \| \mathcal{N}_+ \mathcal{U}_N \psi_N \| \, .
\end{align}
where we used that $\mathcal{N} \mathcal{U}_N \psi_N = \mathcal{N}_+ \mathcal{U}_N \psi_N$. From \eqref{def:h,H} we get
\begin{align}
\label{eq:bound-h}
\| h \|_{L^2_{\perp \varphi} ( \Lambda )}  = \| q \widetilde{g} (O) \varphi \|_{L^2_{\perp \varphi} ( \Lambda )} \leq \| g \|_\infty \| \varphi \|_{L^2( \Lambda)} =  \| g \|_\infty
\end{align}
and
\begin{align}
\label{eq:bound-H}
\| H \| \leq \| q \widetilde{g}(O) q \| \leq \| g \|_\infty
    \, .
\end{align}
All in all, with \eqref{eq:bound-N}, \eqref{eq:bound-h} we arrive at
\begin{align}
\frac{1}{N^{3/2}}   \vert \langle \mathcal{U}_N  \psi_N, \, \left( \Big[ b^*(h) + b(h) \Big] d \Gamma (H) + d \Gamma (H) \Big[ b^*(h) + b(h) \Big] \right) \psi_N \rangle\vert
\leq \frac{4 \| g \|_\infty^2}{N^{3/2}}  \, .
\label{eq:variance-3-2}
\end{align}
For the second term of the r.h.s. of \eqref{eq:variance-4}, using  \eqref{eq:bound-N} and \eqref{eq:bound-H}, we similarly find that
\begin{align}
\frac{1}{N^2} \vert \langle\mathcal{U}_N  \psi_N, \Big[d\Gamma (H) \Big]^2 \mathcal{U}_N \psi_N \rangle \vert  \leq \frac{1}{N^2} \| H \|^2 \| ( \mathcal{N}_+ + 1) \mathcal{U}_N \psi_N \|^2 \leq \frac{\| g \|_\infty^2}{N^2} \, .  \label{eq:variance-3-3}
\end{align}
Thus with \eqref{eq:variance-3-2} and \eqref{eq:variance-3-3} we conclude from \eqref{eq:variance-3} the sought estimate
\begin{align}
\Big\vert \langle \psi_N,  & \, \Big[ \frac{1}{N} \sum_{i=1}^N g(O^{(i)}) - \langle \varphi,g( O) \varphi \rangle \Big]^2 \psi_N \rangle \notag -  \frac{1}{N} \langle\mathcal{U}_N  \psi_N, \Big[ b^*(h) + b(h) \Big]^2 \mathcal{U}_N \psi_N \rangle \Big\vert  \leq C N^{-3/2}
\end{align}
for some $C>0$.
\subsubsection*{Step 2 (Norm approximation):} Next we use \eqref{eq:norm-approx} to approximate the excitation vector $\mathcal{U}_N \psi_N$. Using the property of the Bogoliubov transform, we claim that there exists $C>0$ such that
\begin{align}
\Big\vert \frac{1}{N}  & \langle\mathcal{U}_N  \psi_N, \Big[ b^*(h) + b(h) \Big]^2 \mathcal{U}_N \psi_N \rangle - \frac{1}{N} \langle e^{B(\eta)} e^{-A} e^{B (\tau)} \Omega , \Big[ b^*(h) + b(h) \Big]^2  e^{B(\eta)} e^{-A} e^{B (\tau)} \Omega \rangle \Big\vert \notag \\
\leq& \frac{C}{N^{5/4}} \, .
\end{align}
To this purpose, recall the definition of $\omega \in [0, 2\pi]$ in \eqref{eq:norm-approx}. By  triangle inequality we find 
\begin{align}
\vert \frac{1}{N}  & \langle\mathcal{U}_N  \psi_N, \Big[ b^*(h) + b(h) \Big]^2 \mathcal{U}_N \psi_N \rangle - \frac{1}{N} \langle e^{B(\eta)} e^{-A} e^{B (\tau)} \Omega , \Big[ b^*(h) + b(h) \Big]^2  e^{B(\eta)} e^{-A} e^{B (\tau)} \Omega \rangle \vert \notag \\
\leq& \frac{1}{N} \| e^{i \omega}e^{B(\eta)} e^{-A} e^{B (\tau)} \Omega - \mathcal{U}_N \psi_N \| \, \| \Big[ b^*(h) + b(h) \Big]^2 \mathcal{U}_N \psi_N \| \notag \\
&+ \frac{1}{N} \| e^{i \omega}e^{B(\eta)} e^{-A} e^{B (\tau)} \Omega - \mathcal{U}_N \psi_N \| \, \| \Big[ b^*(h) + b(h) \Big]^2 e^{i \omega}e^{B(\eta)} e^{-A} e^{B (\tau)} \Omega\|  \, .
\end{align}
With \eqref{eq:bounds.b}, \eqref{eq:bound-h} and $b(h) \mathcal{N}_+ = (\mathcal{N}_+ +1) b(h)$ (from the commutation relations), we end up with
\begin{align}
\| \Big[ b^*(h) + b(h) \Big]^2 \mathcal{U}_N \psi_N \| \leq& 2 \| g \|_\infty \| ( \mathcal{N}_+ + 1)^{1/2} (b^*(h) + b(h) ) \mathcal{U}_N \psi_N  \| \notag \\
\leq& 2 \| g \|_\infty \| b^*(h) \mathcal{N}_+^{1/2} \mathcal{U}_N \psi_N  \| + 2 \| g \|_\infty \| b(h) (\mathcal{N}_+ + 2)^{1/2} \mathcal{U}_N \psi_N  \| \,.
\end{align}
The norm approximation \eqref{eq:norm-approx} and  \eqref{eq:bounds.b} thus yield
 \begin{align}
\vert \frac{1}{N}  & \langle\mathcal{U}_N  \psi_N, \Big[ b^*(h) + b(h) \Big]^2 \mathcal{U}_N \psi_N \rangle - \frac{1}{N} \langle e^{B(\eta)} e^{-A} e^{B (\tau)} \Omega , \Big[ b^*(h) + b(h) \Big]^2  e^{B(\eta)} e^{-A} e^{B (\tau)} \Omega \rangle \vert \notag \\
\leq& \frac{C}{N^{5/4}}  \, \| ( \mathcal{N}_+ + 1) \mathcal{U}_N \psi_N \| + \frac{1}{N^{5/4}}  \, \| ( \mathcal{N}_+ + 1) e^{B(\eta)} e^{-A} e^{B (\tau)} \Omega\|  \, .
\end{align}
Furthermore from \eqref{eq:bound-N} and since the number of particles is approximately preserved along $e^B(\eta),e^{B( \tau)}$ resp. $e^A$ (see \eqref{eq:bogo-N}, \eqref{eq:eA-N}, \eqref{eq:bogo-N-2}), i.e.
 \begin{align}
\| ( \mathcal{N}_+ & + 1)  e^{B(\eta)} e^{-A} e^{B (\tau)} \Omega \| \notag \\
&\leq C \|( \mathcal{N}_+ + 1)  e^{-A} e^{B (\tau)} \Omega \| \leq C \|  ( \mathcal{N}_+ + 1)  e^{B (\tau)} \Omega \| \leq C \|  ( \mathcal{N}_+ + 1) \Omega \|
    \, , 
\end{align}
we arrive as $\mathcal{N}_+ \Omega =0$ at
 \begin{align}
\vert \frac{1}{N}  & \langle\mathcal{U}_N  \psi_N, \Big[ b^*(h) + b(h) \Big]^2 \mathcal{U}_N \psi_N \rangle - \frac{1}{N} \langle e^{B(\eta)} e^{-A} e^{B (\tau)} \Omega , \Big[ b^*(h) + b(h) \Big]^2  e^{B(\eta)} e^{-A} e^{B (\tau)} \Omega \rangle \vert \notag \\
\leq& \frac{C}{N^{5/4}}
\end{align}
\subsubsection*{Step 3 (Action of $e^{B(\eta)}$)} 
Next, we show that by the Bogoliubov transform's action \eqref{eq:action-bogo} on modified creation and annihilation operators, it follows there exists $C>0$ such that
\begin{align}
\label{eq:h-h1}
    \Big\vert \frac{1}{N} & \langle e^{B(\eta)} e^{-A} e^{B (\tau)} \Omega , \Big[ b^*(h) + b(h) \Big]^2  e^{B(\eta)} e^{-A} e^{B (\tau)} \Omega \rangle  \\
    & \notag 
    \quad - \frac{1}{N} \langle e^{-A} e^{B (\tau)} \Omega , \Big[ b^*(h_1) + b(h_1) \Big]^2   e^{-A} e^{B (\tau)} \Omega \rangle \Big\vert \leq \frac{C \| g \|_\infty}{N^2} \, , 
\end{align}
 where $h_1(p) =\cosh (\eta_p )h_p  + \sinh (\eta_{p}) h_{-p}$. To this end, we compute with \eqref{eq:action-bogo}
 \begin{align}
 \frac{1}{N} & \langle e^{B(\eta)} e^{-A} e^{B (\tau)} \Omega , \Big[ b^*(h) + b(h) \Big]^2  e^{B(\eta)} e^{-A} e^{B (\tau)} \Omega \rangle \notag \\
 =&\frac{1}{N} \langle e^{-A} e^{B (\tau)} \Omega , \Big[ b^*(h_1) + b(h_1)  + d_\eta (h) + d^*_\eta (h) \Big]^2   e^{-A} e^{B (\tau)} \Omega \rangle
 \end{align}
We will show that the operators $d_\eta (h), d^*_\eta (h)$ contribute subleading only; in fact it follows that
 \begin{align}
    \Big\vert \frac{1}{N} & \langle e^{B(\eta)} e^{-A} e^{B (\tau)} \Omega , \Big[ b^*(h) + b(h) \Big]^2  e^{B(\eta)} e^{-A} e^{B (\tau)} \Omega \rangle 
\\
    &\qquad \qquad \qquad \qquad - \frac{1}{N} \langle e^{-A} e^{B (\tau)} \Omega , \Big[ b^*(h_1) + b(h_1) \Big]^2   e^{-A} e^{B (\tau)} \Omega \rangle \Big\vert \notag \\
    \leq& \frac{1}{N}  \| d_\eta (h)  e^{-B(\eta)}\Big[ b^*(h) + b(h) \Big]  e^{B(\eta)}e^{-A} e^{B (\tau)} \Omega \| \notag \\
    &\quad + \frac{1}{N} \|  e^{-B(\eta)}\Big[ b^*(h) + b(h) \Big]  e^{B(\eta)}e^{-A} e^{B (\tau)} \Omega \| \,  \|  d_\eta (h)  e^{-A} e^{B (\tau)} \Omega  \|  \,  \|   e^{-A} e^{B (\tau)} \Omega  \| \notag \\
    &\quad + \frac{1}{N}  \|  d_\eta (h)  \Big[ b^*(h_1) + b(h_1) \Big] e^{-A} e^{B (\tau)} \Omega \| \notag \\
    &\quad +  \frac{1}{N}  \|  d_\eta (h)  e^{-A} e^{B (\tau)} \Omega \| \, \| \Big[ b^*(h_1) + b(h_1) \Big] e^{-A} e^{B (\tau)} \Omega \| \, . 
    \label{eq:variance-5}
 \end{align}
For the first term of the r.h.s. we find from \eqref{eq:bound-d} and \eqref{eq:bogo-N}
\begin{align}
\| d_\eta (h)   & e^{-B(\eta)}\Big[ b^*(h) + b(h) \Big]  e^{B(\eta)}e^{-A} e^{B (\tau)} \Omega \| \notag \\
\leq&  \frac{C}{N} \| ( \mathcal{N}_+ + 1)^{3/2} e^{-B(\eta)}\Big[ b^*(h) + b(h) \Big]  e^{B(\eta)}e^{-A} e^{B (\tau)} \Omega  \| \notag \\
\leq& \frac{C}{N} \| ( \mathcal{N}_+ + 1)^{3/2} \Big[ b^*(h) + b(h) \Big]  e^{B(\eta)}e^{-A} e^{B (\tau)} \Omega \| \, .
\end{align}
Furthermore with $ \mathcal{N}_+ b^*(f) = b^*(f) ( \mathcal{N}_+ - 1) $ and \eqref{eq:bounds.b}, \eqref{eq:eA-N} and \eqref{eq:bogo-N-2}
 \begin{align}
\| d_\eta (h)  & e^{-B(\eta)}\Big[ b^*(h) + b(h) \Big]  e^{B(\eta)}e^{-A} e^{B (\tau)} \Omega \| \notag \\
\leq&   \frac{C}{N} \| ( \mathcal{N}_+ + 1)^{2} e^{B(\eta)}e^{-A} e^{B (\tau)} \Omega \| \leq  \frac{C}{N} \| ( \mathcal{N}_+ + 1)^{2}  \Omega \| \leq \frac{C}{N} \, .
\end{align}
 We proceed similarly for the remaining three terms of the r.h.s. of \eqref{eq:variance-5} and with
\begin{align}
\| h_1 \|_{\ell^2( \Lambda_+^*)} \leq C \| \eta \|_{\ell^\infty( \Lambda_+^*)} \| h \|_{\ell^2( \Lambda_+^*)} \leq C \| g \|_{\infty}
\end{align}
we conclude the proof of the claimed bound \eqref{eq:h-h1}.
% \begin{align}
% \Big\vert \frac{1}{N} & \langle e^{B(\eta)} e^{-A} e^{B (\tau)} \Omega , \Big[ b^*(h) + b(h) \Big]^2  e^{B(\eta)} e^{-A} e^{B (\tau)} \Omega \rangle \notag \\
% & \quad - \frac{1}{N} \langle e^{-A} e^{B (\tau)} \Omega , \Big[ b^*(h_1) + b(h_1) \Big]^2   e^{-A} e^{B (\tau)} \Omega \rangle \vert \leq \frac{C \| g \|_\infty^2 }{N} \, .
%  \end{align}
 \subsubsection*{Step 4 (Action of $e^A$)}
 Next we show that the unitary $e^A$ contributes subleading only, i.e. that there exists $C>0$  such that
 \begin{align}
  \vert \frac{1}{N}  &\langle e^{-A} e^{B (\tau)} \Omega , \Big[ b^*(h_1) + b(h_1) \Big]^2   e^{-A} e^{B (\tau)} \Omega \rangle - \frac{1}{N} \langle  e^{B (\tau)} \Omega , \Big[ b^*(h_1) + b(h_1) \Big]^2   e^{B (\tau)} \Omega \rangle  \vert \notag \\
 \leq& \frac{C \| g \|_\infty^2}{N^{3/2}} 
   \label{eq:step4}
    \, .
 \end{align}
For that we write
\begin{align}
e^{A} \Big(  b^*(h_1) + b(h_1) \Big)^2 e^{-A}  -
&\Big( b^*(h_1) + b(h_1) \Big)^2
= \int_{-1}^1 ds \, \frac{d}{ds}
e^{sA} \Big( b^*(h_1) + b(h_1) \Big)^2 e^{sA} \notag \\
=&\int_{-1}^1 ds \, e^{sA}\left[ A, \Big( b^*(h_1) + b(h_1) \Big)^2 \right] e^{-sA} \notag \\
=& \int_{-1}^1 ds \, e^{sA}\left[ A,  b^*(h_1) + b(h_1)  \right]\Big( b^*(h_1) + b(h_1) \Big) e^{-sA} \notag \\
&+ \int_{-1}^1 ds \, e^{sA}\Big( b^*(h_1) + b(h_1) \Big) \left[ A,  b^*(h_1) + b(h_1)  \right]\ e^{-sA} \, .
\end{align}
Thus, we obtain that
 \begin{align}
 \vert \frac{1}{N}  &\langle e^{-A} e^{B (\tau)} \Omega , \Big[ b^*(h_1) + b(h_1) \Big]^2   e^{-A} e^{B (\tau)} \Omega \rangle - \frac{1}{N} \langle  e^{B (\tau)} \Omega , \Big[ b^*(h_1) + b(h_1) \Big]^2   e^{B (\tau)} \Omega \rangle  \vert \notag \\
=& \int_{-1}^1 ds \, \vert \langle  e^{B (\tau)} \Omega, \, e^{sA}\left[ A,  b^*(h_1) + b(h_1)  \right]\Big( b^*(h_1) + b(h_1) \Big) e^{-sA}  e^{B (\tau)} \Omega \rangle \vert \notag \\
&+ \int_{-1}^1 ds \, \vert \langle  e^{B (\tau)} \Omega , e^{sA}\Big( b^*(h_1) + b(h_1) \Big) \left[ A,  b^*(h_1) + b(h_1)  \right]\ e^{-sA}  e^{B (\tau)} \Omega \rangle \vert
    \, .
 \end{align}
 We now simly apply \cite[Lemma 3.1]{RS} to find 
 \begin{align}
  \vert \frac{1}{N}  &\langle e^{-A} e^{B (\tau)} \Omega , \Big[ b^*(h_1) + b(h_1) \Big]^2   e^{-A} e^{B (\tau)} \Omega \rangle - \frac{1}{N} \langle  e^{B (\tau)} \Omega , \Big[ b^*(h_1) + b(h_1) \Big]^2   e^{B (\tau)} \Omega \rangle  \vert \notag \\
 \leq& \frac{C \| g \|_\infty^2}{N^{3/2}} \int_{-1}^1 ds \, \| ( \mathcal{N}_+ + 1)^{1/2} e^{sA} e^{B( \tau)} \Omega \| \, \| ( \mathcal{N}_+ + 1)^{1/2}  e^{sA} e^{B (\tau)} \Omega \|
 \end{align}
Furthermore, using again \eqref{eq:eA-N} and \eqref{eq:bogo-N-2}, we conclude the proof of \eqref{eq:step4}.
% \begin{align}
%   \vert \frac{1}{N}  &\langle e^{-A} e^{B (\tau)} \Omega , \Big[ b^*(h_1) + b(h_1) \Big]^2   e^{-A} e^{B (\tau)} \Omega \rangle - \frac{1}{N} \langle  e^{B (\tau)} \Omega , \Big[ b^*(h_1) + b(h_1) \Big]^2   e^{B (\tau)} \Omega \rangle  \vert \notag \\
%  \leq& \frac{C}{N^{3/2}} \, .
%  \end{align}
\subsubsection*{Step 5 (Action of $e^{B(\tau)}$)} Since $\tau$ is bounded in $\ell^2,\ell^\infty$ independent in $N$, and thus for  $h_2 (p) = \cosh (\tau_p) h_1 (p) + \sinh(\tau_p) h_1(p) $ we have
\begin{align}
 \| h_2 \|_{\ell^2( \Lambda_+^*)} \leq C \| h_1 \|_{\ell^2 ( \Lambda_+^*)} \leq C\| g \|_\infty \, .
 \end{align}
Therefore, it follows similarly to Step 3 that there exists $C>0$ such that
\begin{align}
 \vert \frac{1}{N}  &\langle  e^{B (\tau)} \Omega , \Big[ b^*(h_1) + b(h_1) \Big]^2    e^{B (\tau)} \Omega \rangle - \frac{1}{N} \langle \Omega , \Big[ b^*(h_2) + b(h_2) \Big]^2    \Omega \rangle  \vert \notag \\
 \leq& \frac{C \| g \|_\infty^2}{N^{1}} \, .
\end{align}

\subsubsection*{Step 5 (Computation of expectation value)} As a last step, we compute the expectation value
 \begin{align}
 \frac{1}{N}& \langle \Omega , \Big[ b^*(h_2) + b(h_2) \Big]^2    \Omega \rangle \notag \\
 &= \frac{1}{N} \langle \Omega ,  \Big( b^*(h_2) b(h_2) + b^*(h_2)^*b(h_2)+ b(h_2) b(h_2) +b(h_2) b^*(h_2)   \Big)  \Omega \rangle
 \end{align}
 Since $b(h) \Omega =0$, we find by the commutation relations
 \begin{align}
  \frac{1}{N}& \langle \Omega , \Big[ b^*(h_2) + b(h_2) \Big]^2    \Omega \rangle \notag = \frac{1}{N} \langle \Omega ,  \left[b(h_2), b^*(h_2)  \right]  \Omega \rangle = \frac{1}{N} \| h_2 \|_{\ell^2( \Lambda_+^*)}^2 \, .
 \end{align}
Finally note that by definition \eqref{eq:tau}, $\tau$ depends on $N$ and thus we are left with computing the asymptotic limit of $\| h_2 \|_{\ell^2( \Lambda_+^*)}^2$. It follows from \cite[Step 4]{RS} that
\begin{align}
\| h_2 - \sigma \|_{\ell^2( \Lambda_+^*)}^2 \leq \frac{C}{N^2} \| g \|_\infty^2
\end{align}
 yielding the final result.
\end{proof}


\begin{thebibliography}{}

\bibitem{AKT}
M. Ajtai, J. Koml\'os, and G. Tusn\'ady.
On optimal mathings. {\em Combinatorica} 4 (1984), pp. 259--264.

\bibitem{AG}
{L. Ambrosio, N. Gigli}, 
{A user's guide to optimal transport. Modelling and optimisation of
flows on networks}, 
\newblock{\textit{Lecture Notes in Math.}, 2062, Fond. CIME/CIME Found. Subser., Springer, Heidelberg, 2013.}

\bibitem{Wiemann} M. H. Anderson, J. R. Ensher, M. R. Matthews, C. E. Wieman, and E. A. Cornell. 
Observation of Bose-Einstein condensation in a dilute atomic vapor. {\em Science} 269 
(1995), pp. 198--201.


\bibitem{BKS} G. Ben Arous, K. Kirkpatrick, and B. Schlein, A central limit theorem in many-body quantum dynamics, {\em Comm. Math. Phys.}, 321(2):371-417 (2013). 

\bibitem{simple-AKT}
S. Bobkov and M. Ledoux. A simple Fourier analytic proof of the AKT optimal matching theorem. arXiv preprint arXiv:1909.06193

\bibitem{Bobkov-Ledoux}
S. Bobkov and M. Ledoux. One-dimensional empirical measures, order statistics, and Kantorovich transport distances. {\em Memoirs of the Amer. Math. Soc.} 261 (2019).


\bibitem{BBCS_cond} 
C. Boccato, C. Brennecke, S. Cenatiempo and B. Schlein. Complete {B}ose-{E}instein condensation in the {G}ross-{P}itaevskii regime. {\em Commun. Math. Phys.} 359 (2018), pp. 975--1026. 

\bibitem{BBCS} C. Boccato, C. Brennecke, S. Cenatiempo and B. Schlein. Bogoliubov Theory in the {G}ross-{P}itaevskii Limit. {\em Acta Mathematica} 222 (2019), pp.  219--335.

\bibitem{BBCS_optimal} 
C. Boccato, C. Brennecke, S. Cenatiempo and B. Schlein. Optimal Rate for Bose-Einstein Condensation in the Gross-Pitaevskii Regime. {\em Commun. Math. Phys.} 376 (2020), pp. 1311--1395. 

\bibitem{Bose} S. Bose. Plancks Gesetz und Lichtquantenhypothese. {\em Z. Phys.} 26 (1924), pp. 178--181.

\bibitem{BSS}
C. Brennecke, B. Schlein, and S. Schraven. Bogoliubov Theory for Trapped Bosons in the {G}ross-{P}itaevskii Regime. {\em Ann. Henri Poincar\'{e}} 23 (2022), pp. 1583--1658. 

\bibitem{BuSS} S. Buchholz, C. Saffirio, and B. Schlein, Multivariate central limit theorem in quantum dynamics, {\em J. Stat. Phys.}, 154(1-2):113?152 (2014).

\bibitem{COS} C. Caraci, J. Oldenburg, B. Schlein, Quantum Fluctuations of Many-Body Dynamics around the Gross-Pitaevskii Equation, Preprint: arXiv:2308.11687

\bibitem{DM-2017}
J. Dedecker, and F. Merlevede. 
Behavior of the Wasserstein distance between the empirical and the marginal distributions of stationary $\alpha$-dependent sequences. {\em Bernoulli} 23 (2017), 2083--2127.

\bibitem{Einstein} A. Einstein. Quantentheorie des einatomigen idealen Gases.  {\em Sitzungsberichte der
Preu\ss ischen Akademie der Wissenschaften} XXII (1924), pp. 261--267.

\bibitem{Figalli}
A. Figalli, F. Glaudo,
An Invitation to Optimal Transport, Wasserstein Distances, and Gradient Flows,
\textit{EMS Textbooks in Mathematics}, Volume 23, 2021.

\bibitem{founier-guillin}
N. Fournier and A. Guillin.
On the rate of convergence in Wasserstein distance of the empirical measure. {\em Probab. Theory Relat. Fields} 162 (2015), pp. 707--738.

\bibitem{Ketterle}  K. B. Davis, M. O. Mewes, M. R. Andrews, N. J. van Druten, D. S. Durfee, D. M.
Kurn, and W. Ketterle. Bose-Einstein Condensation in a Gas of Sodium Atoms. {\em Phys.
Rev. Lett.} 75 (1995), pp. 3969--3973.


\bibitem{KRS} K. Kirkpatrick, S. Rademacher, B. Schlein. A large deviation principle in many-body quantum dynamics. {\em Ann. Henri Poincare}, 22, 2595-2618 (2021).


\bibitem{LNSS} 
M. Lewin, P.T. Nam, S. Serfaty and J.P. Solovej. Bogoliubov spectrum of
interacting Bose gases. {\em Comm. Pure Appl. Math.} 68 (2014), pp. 413--471.

\bibitem{LS_cond} 
E.H. Lieb and R. Seiringer. Proof of Bose-Einstein Condensation for Dilute Trapped Gases. {\em Phys. Rev. Lett.} 88 (2002), p. 170409.


\bibitem{LS} 
E.H. Lieb, and R. Seiringer, Derivation of the {G}ross-{P}itaevskii equation for rotating Bose gases. {\em Commun. Math. Phys.} 264 (2006), pp. 505--537. 

\bibitem{NNRT} 
P.T. Nam, M. Napi\'{o}rkowski, J. Ricaud, and A. Triay. Optimal rate of condensation for trapped bosons in the Gross-Pitaevskii regime. {\em Analysis \& PDE} 15 (2022), pp.  1585--1616. 


\bibitem{NR} P.T. Nam, S. Rademacher, Exponential bounds of the condensation for dilute Bose gases, Preprint: arXiv:2307.10622.

\bibitem{NRS} 
P.T. Nam, N. Rougerie and  R. Seiringer.  Ground states of large bosonic systems: The {G}ross-{P}itaevskii limit revisited. {\em Analysis \& PDE}  9 (2016), pp. 459--485.


\bibitem{NT} P.T. Nam and  A. Triay. {B}ogoliubov excitation spectrum of trapped Bose gases in the {G}ross-{P}itaevskii regime. {\em J. Math. Pures Appl.}, to appear (arXiv:2106.11949). 


\bibitem{RS} S. Rademacher, B. Schlein, Central limit theorem for Bose-Einstein condensates, J. Math. Phys. 60, 071902 (2019). 

\bibitem{R} S. Rademacher, Dependent random variables in quantum dynamics, J. Math. Phys, 60 (8), 2022. 

\bibitem{R-LD2023} S. Rademacher. Large deviations for the ground state of weakly interacting Bose gases. arXiv preprint arXiv:2301.00430 

\bibitem{R-2020} S. Rademacher.
Central limit theorem for Bose gases interacting through singular potentials. {\em Lett. Math. Phys.} 110 (2020), pp. 2143--2147.

\bibitem{RS-JSP-22} S. Rademacher and R. Seiringer. Large deviation estimates for weakly interacting bosons. {\em J. Stat. Phys} 188 (2022), article no. 9.

\bibitem{Santambrogio}
{F. Santambrogio}, 
Optimal Transport for Applied Mathematicians, 
\textit{Progress in Nonlinear Differential
Equations and Their Applications} 87, Birkhauser Basel (2015).

\bibitem{V} C. Villani, Topics in Optimal Transportation, \textit{Graduate Studies in Mathematics.} vol. 58 of, Amer. Math. Soc. Providence, RI, 2003. 		
		

\bibitem{Villani} 
C. Villani, Optimal Transport: Old and New,
\textit{Grundlehren der mathematischen Wissenschaften} 338, Springer-Verlag Berlin Heidelberg, 2009.

\end{thebibliography}
\end{document}